\documentclass[natbib]{svjour3}     
\smartqed  

\usepackage{graphicx}
\usepackage{amsbsy}
\usepackage{epsfig,rotating,graphicx}
\usepackage{latexsym,amsmath,amssymb}
\usepackage{amssymb}   
\usepackage{url}    

\newcommand{\vel}{\mbox{{\boldmath$v$}}} 
\newcommand{\id}{\mbox{$\rm d$}} 
\newcommand{\br}{\mbox{{\boldmath$r$}}} 

\begin{document}

\title{Helioseismology of Sunspots: A Case Study of NOAA Region 9787}

\titlerunning{Helioseismology of Sunspots}   


\author{L. Gizon  \and   H. Schunker \and C.S. Baldner \and S. Basu \and A.C. Birch \and  R.S. Bogart  \and D.C. Braun \and R. Cameron \and  T.L. Duvall Jr. \and S.M. Hanasoge \and J. Jackiewicz \and  M. Roth \and T. Stahn  \and M.J. Thompson \and S. Zharkov}

\authorrunning{Gizon et al.} 

\institute{L. Gizon \and H. Schunker \and R. Cameron \and J. Jackiewicz \and M. Roth \and T. Stahn \at
              Max-Planck-Institut f{\"u}r Sonnensystemforschung, 37191 Katlenburg-Lindau, Germany\\ 
 \email{gizon@mps.mpg.de}           
          \and
           C.S. Baldner \and S. Basu \at Department of Astronomy, Yale University, PO Box 208101, New Haven, CT 06520, USA 
\and
           A.C. Birch \and D.C. Braun \at
Colorado Research Associates, a division of NorthWest Research Associates, Inc., 3380 Mitchell Lane, Boulder, CO 80301-5410, USA 
         \and
         R.S. Bogart \and S.M. Hanasoge \at Hansen Experimental Physics Laboratory, Stanford University, Stanford, CA 94305, USA 
         \and
         T.L. Duvall Jr. \at
         Laboratory for Solar Physics, NASA/Goddard Space Flight Center, Greenbelt, MD 20771, USA
         \and
         M.J. Thompson \and S. Zharkov \at
         School of Mathematics and Statistics, University of Sheffield, Houndsfield Road, Sheffield, S3 7RH, UK 
}

\date{Received: date / Accepted: date}

\maketitle

\begin{abstract}
Various methods of helioseismology are used to study the subsurface properties of the sunspot in NOAA Active Region~9787. This sunspot was chosen because it is axisymmetric, shows little evolution during 20-28 January 2002, and was observed continuously by the MDI/SOHO instrument. AR~9787 is visible on helioseismic maps of the farside of the Sun from 15 January, {i.e.} days before it crossed the East limb.

Oscillations have reduced amplitudes in the sunspot at all frequencies, whereas a region of enhanced acoustic power above 5.5 mHz (above the quiet-Sun acoustic cutoff) is seen outside the sunspot and the plage region. This enhanced acoustic power has been suggested to be caused by the conversion of acoustic waves into magneto-acoustic waves that are refracted back into the interior and re-emerge as acoustic waves in the quiet Sun. 
Observations show that the sunspot absorbs a significant fraction of the incoming $p$ and $f$ modes around 3 mHz.  A numerical simulation of MHD wave propagation through a simple model of AR~9787 confirmed that wave absorption is likely to be due to the partial conversion of incoming waves into magneto-acoustic waves that propagate down the sunspot. 

Wave travel times and mode frequencies are affected by the sunspot. In most cases, wave packets that propagate through the sunspot have reduced travel times. At short travel distances, however, the sign of the travel-time shifts appears to depend sensitively on how the data are processed and, in particular, on filtering in frequency-wavenumber space.  We carry out two linear inversions for wave speed: one using travel-times and phase-speed filters and the other one using mode frequencies from ring analysis. These two inversions give subsurface wave-speed profiles with opposite signs and different amplitudes. 

The travel-time measurements also imply different subsurface flow patterns in the surface layer depending on the filtering procedure that is used. Current sensitivity kernels are unable to reconcile these measurements, perhaps because they rely on imperfect models of the power spectrum of solar oscillations. We present a linear inversion for flows of ridge-filtered travel times. This inversion shows a horizontal outflow in the upper 4~Mm that is consistent with the moat flow deduced from the surface motion of moving magnetic features. 

From this study of AR~9787, we conclude that we are currently unable to provide a unified description of the subsurface structure and dynamics of the sunspot.  
\end{abstract}


\section{Introduction}
\label{intro}

One of the main goals of solar physics is to understand the physical processes responsible for solar magnetism and activity. This requires the study of magnetic flux tubes, their transport and  dynamics in the convection zone, and their emergence at the solar surface in the form of  sunspots and active regions. The overall nature of sunspots is still a matter of debate. Many open questions remain concerning their structure and, above all, their formation and stability.  How can regions of such intense magnetic flux come into existence and remain stable over several days, weeks, and sometimes months?   Another common question is whether sunspots are monolithic magnetic flux tubes or have a spaghetti-like structure \citep{Parker1979}. \citet{Schuessler2005} proposed a scenario whereby a sunspot expands rapidly below the surface during the early stages of its formation, leading to a disconnection from its magnetic roots. This disconnection may allow a transition to a spaghetti-like subsurface structure. As to the stability of sunspots, it may be due to the presence of surface and subsurface collar flows \citep{Parker1979}. Other questions concern the energetics of sunspots, the flow of heat through and around sunspots, and the nature of magnetoconvection at kilogauss fields.  What is known about sunspots has been summarized by, {e.g.}, \citet{Thomas1992} and \citet{Solanki2003}.  For a description of various magnetostatic sunspot models, we refer the reader to \citet{Jahn1992} and \citet{Rempel2008}. 

In this paper we will discuss the potential of helioseismology to probe the subsurface structure of sunspots, with the hope of answering, one day, some of the questions listed above. Local helioseismology includes several methods of analysis, which have been described in some detail by \citet{Gizon2005}. All these methods rely on continuous time series of Doppler images of the Sun's surface. Fourier-Hankel analysis was developed to study the relationship between ingoing and outgoing waves around a sunspot \citep{Braun1987, Braun1995}.   Ring-diagram analysis consists of analysing the frequencies of solar acoustic waves over small patches of the solar surface \citep{Hill1988, Antia2007}. Time-distance helioseismology (TD) measures the travel times of wave packets moving through the solar interior \citep{Duvall1993}. Helioseismic holography (HH) uses the observed wave field at the solar surface to infer the wave field at different depths \citep{Lindsey1997}. A summary of recent results is provided by \citet{Gizon2006b} and \citet{Thompson2008}.

There have been several studies of sunspots using helioseismology. \citet{Braun1987,Braun1992a} and \citet{Braun1995} used Fourier-Hankel decomposition to measure wave absorption and scattering phase shifts caused by sunspots. The absorption is believed to be the result of a partial conversion of incoming $p$ modes into slow magnetoacoustic waves \citep[e.g.,][]{Spruit1992, Cally2000}.  Observational signatures of the mode conversion process have been discussed, for example, by \citet{Schunker2006b}. Agreement between the observations of \citet{Braun1995} and simplified sunspot models were reported by \citet{Fan1995}, \citet{Cally2003}, and \citet{Crouch2005} using a forward modeling approach. 
Time-distance helioseismology and helioseismic holography aim at making images of the solar interior from maps of travel times or phase shifts under the traditional assumption that the Sun is weakly inhomogeneous in the horizontal directions. TD and HH have been used to infer wave speed variations and flows in and around sunspots \citep[{e.g.}][]{Duvall1996, Jensen2001, Braun2000, Gizon2000, Kosovichev2000, Zhao2001, Couvidat2006}. Ring-diagram analysis has a coarser horizontal resolution and is used to study the subsurface structure of entire active regions \citep[{e.g.}][]{Basu2004, Antia2007,Bogart2008}. While these methods and their variants appear to be quite robust, it has not been demonstrated that they are consistent. For instance ring-diagram analysis has not been directly compared to time-distance or holography in the case of an isolated sunspot. This paper reports on a joint study of the sunspot in NOAA Active Region~9787. 

\section{Observations of NOAA Region 9787}

\subsection{MDI/SOHO Observations} 
\label{ss:thedata}

\begin{figure}
\hspace{-0.5cm}
\includegraphics[width=13.cm]{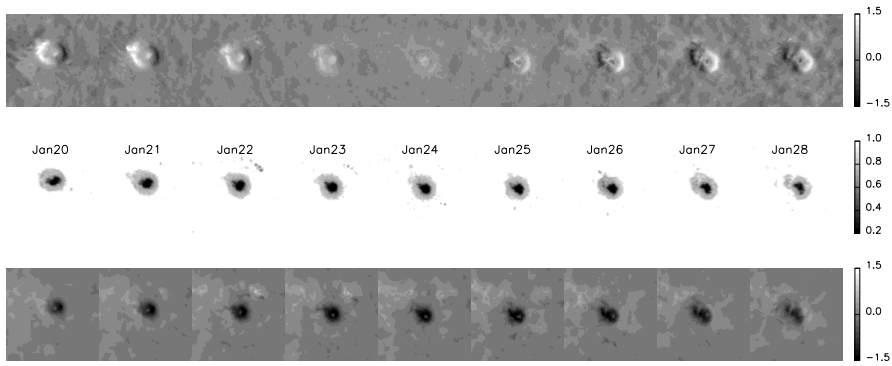}
\caption{Daily averages of the SOHO/MDI Doppler velocity (top), intensity (middle), and line-of-sight magnetic field (bottom) of the sunspot in Active Region~9787 during January 20-28, 2002. The Doppler velocity is in units of km/s, the magnetic field in units of kG.  Each daily frame is a square with sides of length $200$~Mm.
}
\label{evol}
\end{figure}

NOAA Active Region~9787 was chosen from the MDI/SOHO data library because it hosts a large, round, isolated sunspot. A quick look at the data is given by Figure~\ref{evol}.  The data consist of nine days of MDI full disk Dopplergrams for each minute from 20 January to 28 January 2002. MDI also recorded the line-of-sight magnetic field every minute and intensity images every six hours.   The images were remapped using Postel projection with a map scale of $0.12^\circ$. The centers of projection were chosen to track the motion of the sunspot (Carrington longitude $\phi\sim133^\circ$ and  latitude $\lambda=-8.3^\circ$). The re-mapping routine employs a cubic convolution interpolation. Missing data was linearly interpolated in time and a daily temporal mean was subtracted from each Dopplergram. Finally, we are left with one $512 \times 512 \times 1440$ data cube of Doppler velocity data for each day. These data sets are made available on the European Helio- and Asteroseismology Network (HELAS) web site at \url{http://www.mps.mpg.de/projects/seismo/NA4/}. All authors were invited to analyse the same data, thereby eliminating discrepancies in the data reduction methods.
 
\begin{figure}
\begin{center}
\includegraphics[width=12cm]{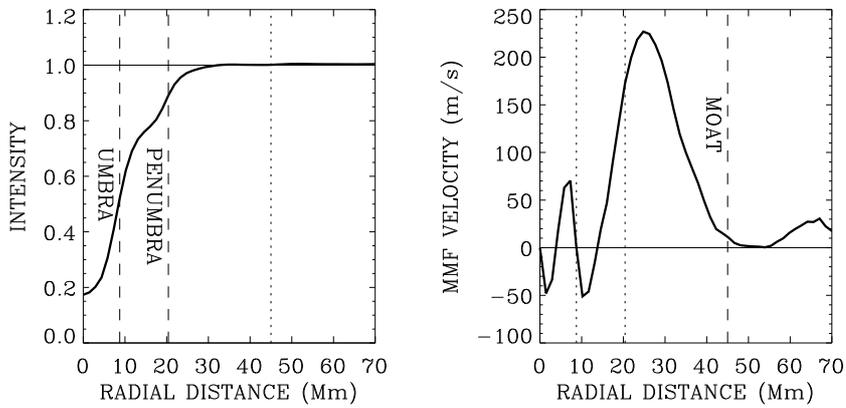}
\caption{(Left) Normalised intensity profile of the sunspot averaged over time and azimuth. The vertical dashed lines indicate the umbral and penumbral boundaries. The boundary of the moat is given by the dotted line. (Right) Velocity of moving magnetic features (MMFs) averaged over time and azimuth as a function of distance from the center of the sunspot. The MMFs track the moat flow and are moving outward from the outer penumbra up to a radius of about 45~Mm (dashed line). The dotted indicate the boundaries of the umbra and the penumbra. }
\label{spot_boundaries}
\end{center}
\end{figure}

  Figure~\ref{evol} shows a daily average of the MDI intensity continuum, magnetic field and Doppler velocity showing that there is little evolution of the sunspot during the period covered by the observations. The Dopplergrams show a $\sim 2$~km/s Evershed outflow in the penumbra of the sunspot. The sunspot exhibits some amount of proper motion.  Figure~\ref{spot_boundaries} shows the intensity profile of the sunspot averaged over nine days and over azimuthal angle, after correcting for the proper motion of the sunspot. The umbral and penumbral boundaries are at radii 9~Mm and 20~Mm respectively.  

The sunspot is surrounded by a region of horizontal outflow called the moat flow. In order to characterize the strength and extent of the moat, we measured the motion of the moving magnetic features (MMFs) from hourly averages of the magnetograms using a local correlation tracking method.  The temporal and azimuthal averages of the MMF velocity is plotted in Figure~\ref{spot_boundaries} as a function of distance from the center of the sunspot. The moat flow has a peak amplitude of $230$~m/s and extends to about $45$~Mm. The moat radius is about twice the penumbral radius, which is a standard value \citep{Brickhouse1988}.

\begin{figure}
\begin{center}
\includegraphics[width=10cm]{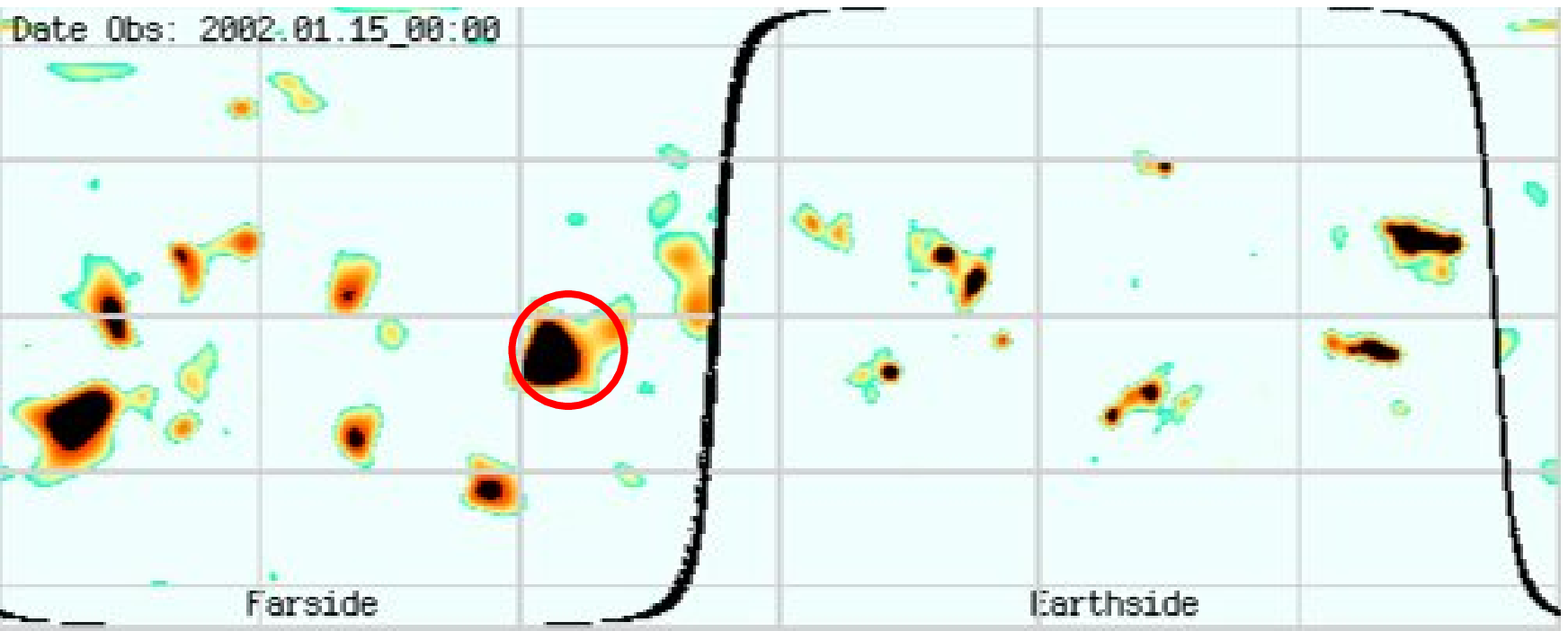}
\includegraphics[width=10cm]{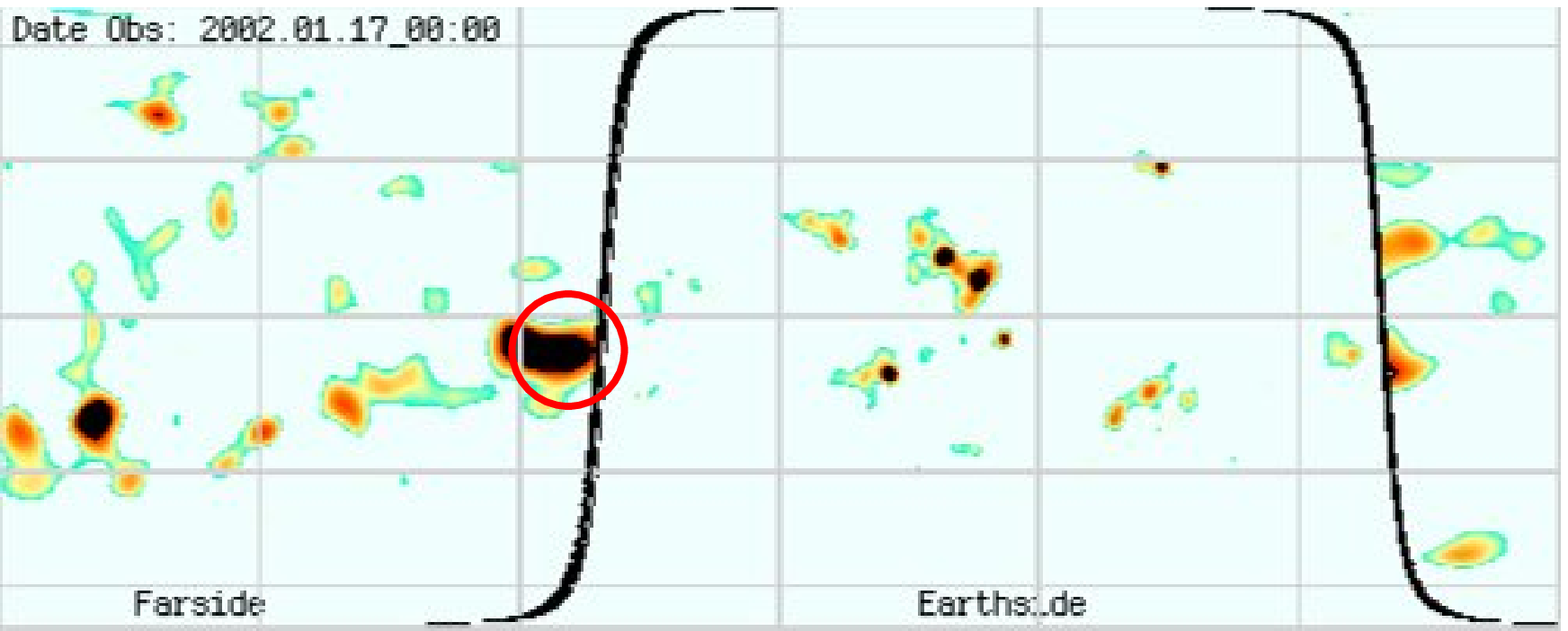}
\includegraphics[width=10cm]{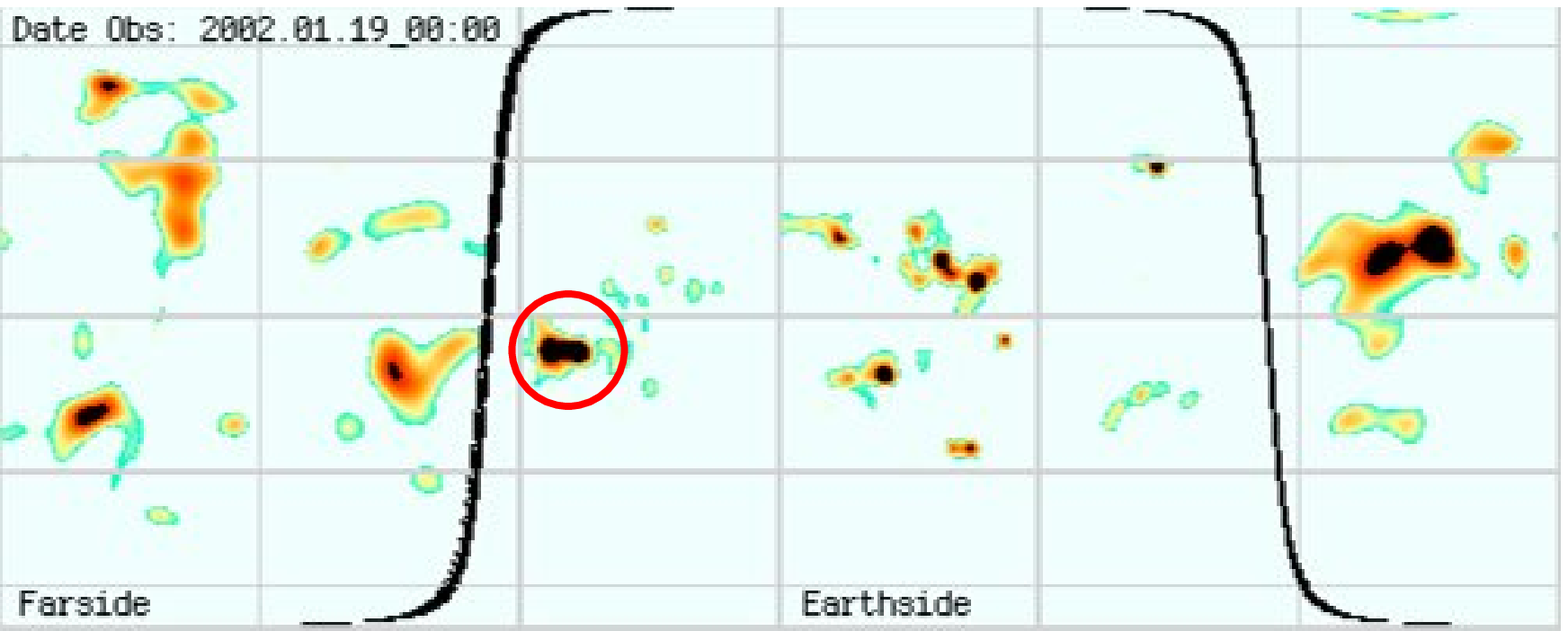}
\includegraphics[width=10cm]{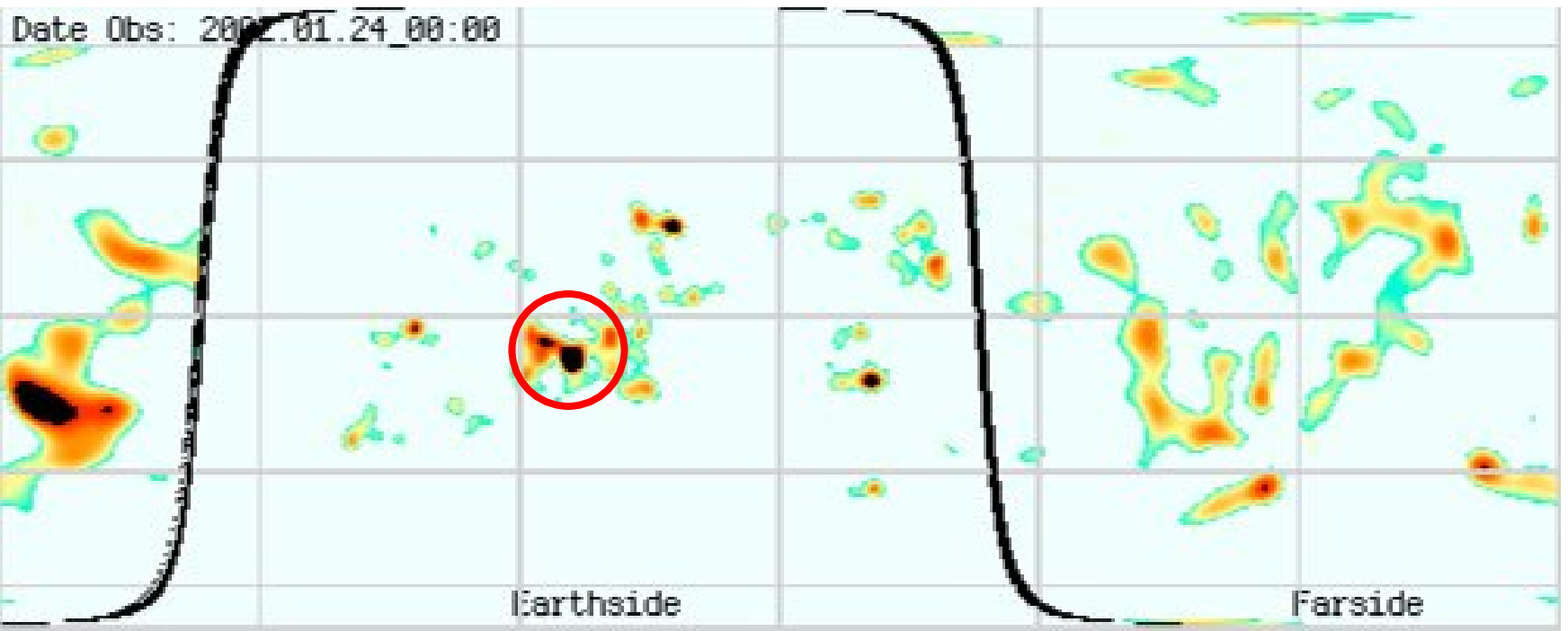}
\end{center}
\caption{Maps of the full Sun created using helioseismic waves to infer the presence of magnetic activity on the farside of the Sun. Strong shifts (black/orange) in the phase indicate an active region, the light blue represents the quiet Sun. The horizontal lines are lines of constant latitude, the vertical lines are lines of constant Carrington longitude separated by $60^\circ$. 
The dates from top to bottom are 2002 June 15, 17, 19, and 24. Active Region~9787 is located close to latitude $\lambda = -8.3^\circ$ and longitude $\phi = 133^\circ$ in all four maps, indicated by the red circles. 
(Courtesy of K. Oslund and P.H. Scherrer (2006))
}
\label{fig:farside}
\end{figure}

The early development of Active Region~9787 can be traced using the helioseismic technique of farside imaging \citep{Braun2000,Lindsey:2000p103}. Figure~\ref{fig:farside}
 shows four maps of the full Sun created by K. Oslund and P.H. Scherrer (2006).
These maps were taken from the SOI website at \url{http://soi.stanford.edu/data/full_farside/}.  The overlaid grid represents Carrington solar coordinates where vertical lines of longitude are separated by $60^\circ$. Active Region~9787 is located close to latitude $\lambda = -8.3^\circ$ and longitude $\phi = 133^\circ$, as shown by the red circles in Figure~\ref{fig:farside}.  The active region is detected on the farside of the Sun in the top two panels of Figure~\ref{fig:farside}. We then see the active region rotate past the East limb to the Earth side (third panel) and then within our observation period (24th January 2002) in the bottom panel.
Additional information is available online at \url{http://news-service.stanford.edu/pr/2006/pr-sun-031506.html}.

\subsection{Oscillatory Power and Acoustic Halos}

There are two main properties of the acoustic power in and around active regions that are well documented. One is the power reduction in strong magnetic field regions, particularly sunspots, and the second is the enhancement of power in the higher frequencies (5--6 mHz)  in the nearby photosphere, known as the acoustic halo \citep{Braun1992b, Donea2000}.  Here we quantify these properties for AR~9787.

\begin{figure}
  \begin{center}
\vspace{-2.cm}
\includegraphics[width=22pc, angle=90]{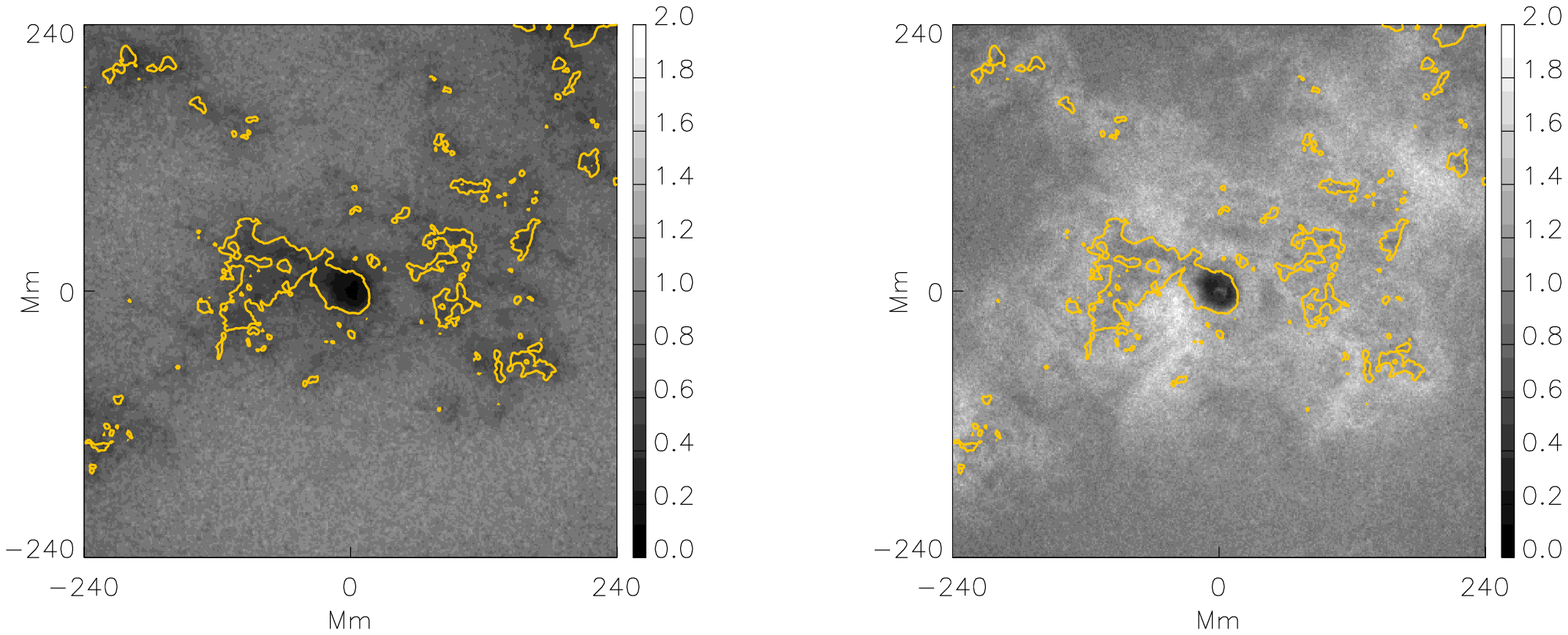}
 \vspace{-2.5cm}
  \caption{Acoustic power averaged over all days and frequency bands $3$--$3.5$~mHz (left) and $5.5$--$6$~mHz (right). Overplotted contour is for $B = 100$~G,  outlining regions of plage. The excess power outside the plage regions is clearly seen in the high-frequency maps (right). The power is normalised to unity in the quiet Sun (grey scale). }
  \label{acpow_maps}
  \end{center}
  \end{figure}

We calculate the temporal Fourier transform of the Doppler images for each day of observation. We divide this into $0.5$~mHz bandwidths and calculate the power averaged over each of these frequency bandwidths. For all frequency bands the acoustic power suppression is greater than $80$\% in the umbra compared to the quiet Sun. We observe enhanced acoustic power at higher frequencies (5--6 mHz) in regions outside the sunspot and strong plage (Figure~\ref{acpow_maps}).

Previous analysis of MDI acoustic power maps by \citet{Ladenkov2002} showed a modest excess of power around a sunspot in the higher frequencies, but which also appeared to be directly related to the location of plage, rather than the sunspot itself.   \citet{Hindman1998} also find that the high frequency velocity signal in an active region is higher (up to 60\%) in pixels with moderate magnetic field strengths between $50$ to $250$~G. The fact that AR~9787 has an extended plage region offers an opportunity to analyze the plage far from the associated sunspot. In essence, this study bolsters the previous analysis of \citet{Hindman1998} and \citet{Ladenkov2002}. 

When the sunspot is located close to the limb we find a significant
enhancement of power in the umbra at high frequencies, to a level almost as high as in the quiet Sun. 
If this power is real and not an artifact of
observing conditions at the limbs, then it could be due to 
 magnetoacoustic waves with a component of motion perpendicular to the field lines in the umbra.

From Figure~\ref{acpow_maps}, a region of particularly strong power close to the south-east side of the sunspot can be seen. This enhanced power is associated with the strong plage region just to the North. \citet{Braun1995} finds evidence of an acoustic deficit immediately outside a small (8~Mm) sunspot, extending out to 35 Mm and appearing to be well defined by the location of surrounding  plage. The sunspot in AR~9787 shows little evidence for a well defined, axisymmetric acoustic halo, leading to the suggestion that the enhanced acoustic power is associated with the strong surrounding plage regions, rather than the sunspot itself.

The results of recent numerical work by \citet{Hanasoge2008} seem to reproduce the high frequency
halo surrounding a small model sunspot. The halo appears between 5 and
6~mHz, close to the acoustic cut-off. In agreement with the suggestion
of \citet{Donea2000}, the upwardly propagating fast
mode waves may be reflected in magnetic regions due to the rapidly
increasing Alfv\'en velocity.  These waves later re-emerge in the region
surrounding the sunspot causing the observed enhanced power.

\subsection{Wave Absorption}
\label{roth}

A useful analysis procedure for studying the interaction of
p modes with sunspots is the decomposition of solar oscillations, 
observed in an annular region around the sunspot, into inward
and outward propagating waves. 
Fourier-Hankel spectral decomposition
has been used to identify $p$-mode absorption in sunspots
and active regions by comparing the amplitudes of the outward and inward
moving waves 
\citep{Braun1988,Bogdan1993,Braun1995}.

Here we use a sunspot-centered spherical-polar coordinate 
system $(\theta,\phi)$ with the sunspot axis at $\theta=0$. 
The annular region is defined by the inner and outer circles at 
$\theta_{\rm min} = 2.5^\circ$ and $\theta_{\rm max}=11.25^\circ$, which correspond to distances between 30 and 137~Mm. The Doppler signal $\psi(\theta,\phi,t)$ in the annular region is decomposed into components of the form
\begin{equation}
{\rm e}^{i(m\phi+2\pi\nu t)}\left[A_m(l,\nu)
H_m^{(1)}(l\theta)+ B_m(l,\nu) H_m^{(2)}(l\theta)\right] ,
\label{eq1}
\end{equation}
where $m$ is the azimuthal order, $l$ is the harmonic degree,
$H_m^{(1)}$ and $H_m^{(2)}$ are Hankel functions of the first and
second kinds, $t$ is time, $\nu$ is temporal frequency, and $A_m$ and
$B_m$ are the complex amplitudes of the incoming and outgoing waves
respectively.  The range in $l$ is between 70 and 1500.
The boundaries of the annulus, $\theta_{\rm min}$ and
$\theta_{\rm max}$, were selected such as to resolve the
low-order $p$ mode ridges with a  resolution in $l$ of approximately 40.
The numerical procedure needed to compute the wave
amplitudes $A_m(l,\nu)$ and $B_m(l,\nu)$ is described
by~\citet{Braun1988}.

\begin{figure}
\hspace{-0.5cm}
\includegraphics[width=12cm]{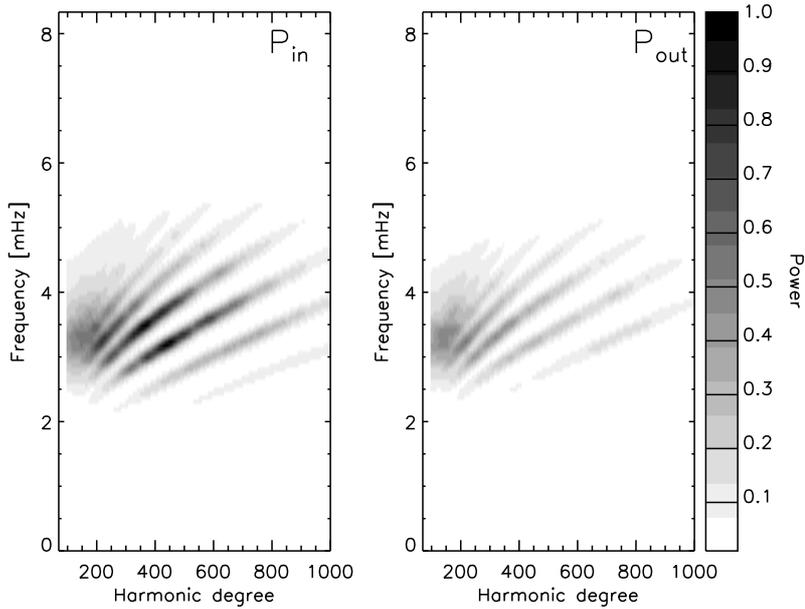}
\caption{Power spectra of inward (left) and outward (right)
propagating waves as a function of harmonic degree  and
frequency. The reduction in power of the outgoing modes is easily noticeable.}
\label{fig1}
\end{figure}

For each value of $l$ we measure the mode amplitudes for the azimuthal
order $m=0$. The power spectra of the incoming and outgoing modes are
displayed in Figure~\ref{fig1}. The outgoing $p$-mode power appears to be
significantly reduced compared to the incoming $p$-mode power.

To measure this more quantitatively we determine an absorption
coefficient for the $f$ and $p_n$ ridges in a 
frequency band between $2.9$ and $3.1$~mHz. 
The absorption coefficient  is defined as
\begin{equation}
\alpha_n =\frac{\int {\id l \, \id \nu} ~W_n (P_{\rm in}-P_{\rm out})}  
{\int {\id l \, \id \nu} ~W_n P_{\rm in} } ,
\end{equation}
where $W_n$ is a window function that selects the $n$-th ridge, and $P_{\rm in}(l,\nu)$ and $P_{\rm out}(l,\nu)$ are the power of the ingoing
and outgoing waves.  
In this frequency band $2.9$--$3.1$~mHz, the $f$, $p_1$, $p_2$, $p_3$, and $p_4$ ridges show absorption coefficients of $57$\%,  $54$\%,  $51$\%,  $49$\%,  and $50$\% respectively. These values are  in agreement with those given in earlier studies \citep{Braun1988,Braun1995} and confirm that AR~9787 ``absorbs'' acoustic waves.

\section{Travel Time Measurements} \label{doug}

\subsection{Phase-Speed Filtering versus Ridge Filtering}
\label{sec.psfilt}

The principle of Helioseismic holography (HH) is to computationally
regress the acoustic amplitudes observed at the surface into
the solar interior \citep{Lindsey1997}.
To facilitate comparisons with results from time-distance analyses,
we use
{\it surface-focused} HH. In the ``space-frequency'' domain, {i.e.}
where $\psi ({\bf r}, \nu)$ denotes the
temporal Fourier transform of the observed Doppler velocities,
the regressions
in surface-focused HH are computed from 
\begin{equation}
H_{\pm}^{P}({\bf r}, \nu) =
\int_{P} \id^2{\bf r}'  
~G_{\pm}({\bf r}, {\bf r}', \nu) ~\psi({\bf r}', \nu ).
\end{equation}
$H_+$ and $H_-$ are the egression and ingression 
which represent estimates of the amplitudes propagating into and out of
the focal point at position ${\bf r}$ on the surface and $\nu$ is the temporal
frequency. $G_+$ and $G_-$ are Green's
functions that express how a monochromatic
point disturbance at a position ${\bf r}'$ on the surface
propagates backward and forward
in time, into the solar interior and back up to the focus. 
They are computed using the
eikonal approximation \citep{Lindsey1997}. 
The correlations,
\begin{equation}
C_{+}^{P}( \textbf{r}) =  \langle {H}_{+}^{P}(\textbf{r}, \nu)
\psi^* (\textbf{r}, \nu)\rangle_{\Delta \nu},
\label{Eq-corr-out}
\end{equation}
and 
\begin{equation}
C_{-}^{P}( \textbf{r}) =  \langle \psi (\textbf{r}, \nu)
{H}_{-}^{P*} (\textbf{r}, \nu) \rangle_{\Delta \nu},
\label{Eq-corr-in}
\end{equation}
describe the egression and ingression control correlations respectively,
which are directly comparable to center-annulus correlations used in 
time-distance (TD) helioseismology \citep[{e.g.}][]{Duvall1996, Braun1997}.
The asterisk denotes complex conjugation, and the brackets indicate
an average over a chosen positive frequency range  $\Delta \nu$.

Surface-focused HH can be used to study flows by dividing
the pupil $P$, over which the ingressions and egressions are computed,
into four quadrants (labeled $N$, $S$, $E$, and $W$), each spanning $90^\circ$ and oriented in the North, South, East and West directions respectively.
We then compute the eight control correlations, $C_{\pm}^{N,S,E,W}$.
Various combinations of these correlations are used to derive
travel-time shifts due to the presence of flows or wave speed
perturbations.  In general, we compute travel-time shifts from
various sums or differences of correlations such that, if
$C$ denotes some linear combination of correlations, the travel-time
shift is 
\begin{equation}
\delta \tau (\textbf{r}) = \arg [ C (\textbf{r}) ] / 2\pi{\nu}_0,
\label{Eq-corr_phase}
\end{equation}
where ${\nu}_0$ is the central frequency of the bandpass $\Delta \nu$.
These represent travel-time shifts of the observed combination of
waves relative to the
travel times expected for
the same ensemble of waves propagating in the solar model used
to compute the Green's functions. 

We present measurements of the mean travel-time shift ($\delta \tau_{\rm mean}$), which represents
the shift computed from the sum of all eight correlations:
\begin{eqnarray}
\delta \tau_{\rm mean} (\textbf{r}) & = &\arg [ 
  C_{+}^{E} (\textbf{r}) + C_{-}^{E} (\textbf{r}) + C_{+}^{W} (\textbf{r}) + 
  C_{-}^{W} (\textbf{r}) + \nonumber \\
& &   C_{+}^{N} (\textbf{r}) + C_{-}^{N} (\textbf{r}) + 
  C_{+}^{S} (\textbf{r}) + C_{-}^{S} (\textbf{r}) ]  / 2\pi{\nu}_0.
\label{Eq-mean}
\end{eqnarray}
We also present travel-time shifts sensitive to horizontal flows. For example,
we define a ``EW'' travel-time asymmetry 
\begin{equation}
\delta \tau_{\rm EW} (\textbf{r}, \nu) = ( \arg [ C_{-}^{E} (\textbf{r}, \nu) + C_{+}^{W} (\textbf{r}, \nu)] 
- \arg [ C_{+}^{E} (\textbf{r}, \nu) + C_{-}^{W} (\textbf{r}, \nu)] ) / 2\pi{\nu}_0.
\label{Eq-EW}
\end{equation} 
A similar travel-time shift can be measured
for a North\,--\,South asymmetry.
The sign of the travel-time perturbations is such that 
a perturbation in the background wave speed resulting in
a faster propagation time will lead to a negative value of the
mean travel-time shift ($\delta \tau_{\rm mean}$) and a horizontal flow directed 
from East to West produces a negative value of the 
EW travel-time asymmetry ($\delta \tau_{\rm EW}$).

Starting with the tracked, Postel-projected datacube described earlier  
(Section~\ref{ss:thedata}) we perform the following steps: 1) a temporal
detrending by subtraction of a linear fit to each pixel signal in time,
2) removal of poor quality pixels, identified by a five-sigma deviation of
any pixel from the linear trend, 3) Fourier transform of
the data in time, 4) a correction
for the amplitude suppression in
magnetic regions \citep{Rajaguru2006},
5) spatial Fourier transform of the data and
multiplication by a chosen filter,
6) extraction of the desired frequency
bandpass, 7) computation of Green's
functions over the appropriate pupils, 8) computation
of ingression and egression amplitudes by a 3D convolution
of the data with the Green's functions, and
9) computation of the travel-time shift maps by
Equations~(\ref{Eq-corr-out})\,--\,(\ref{Eq-EW}).

\begin{figure}[]
\centerline{\scalebox{0.8}{\includegraphics{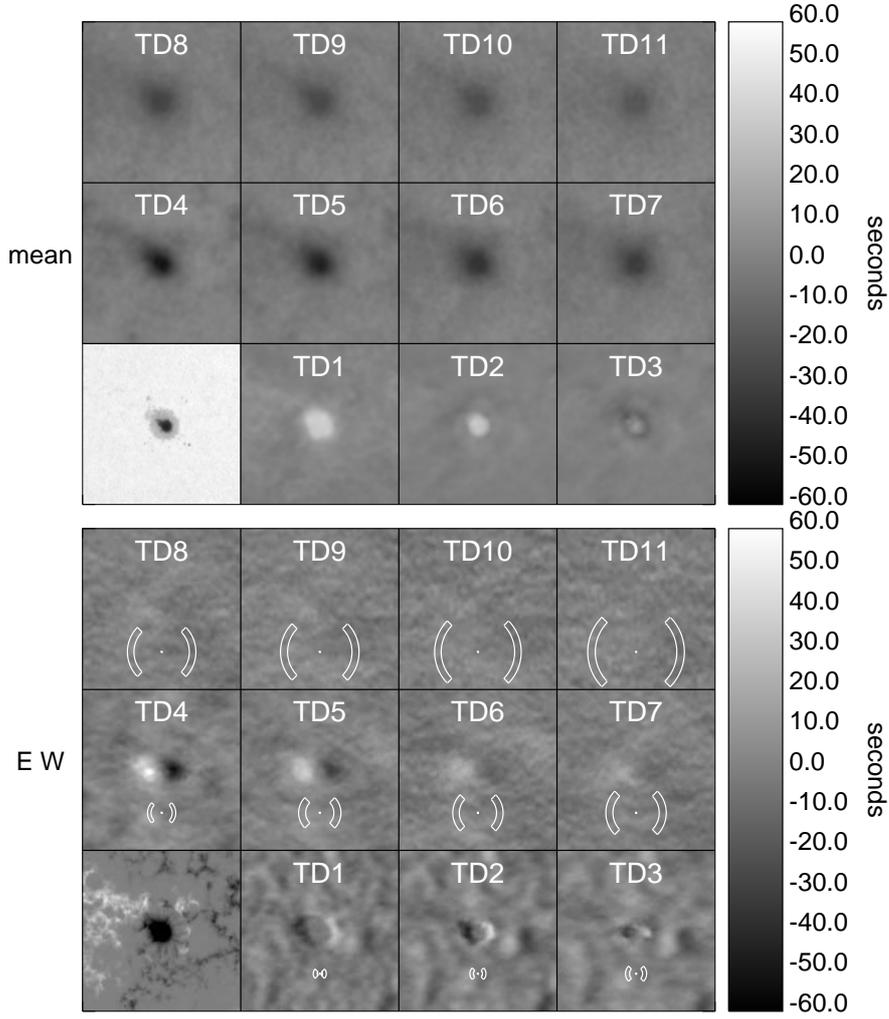}}}
\caption{
Maps of travel-time shifts $\delta \tau_{\rm mean}$ (top panels) and 
$\delta \tau_{\rm EW}$ (bottom panels) using phase speed filters
over a frequency bandpass of 2.5\,--\,5.5 mHz for AR 9787 observed
over a 24 hour period on 2002 January 24.
The labels TD1 through TD11 
indicate the phase-speed filter used. Sizes of the East and West pupil
quadrants used to measure $\delta \tau_{\rm EW}$ are shown
in the bottom set of panels.
The map in the lowest-left position of the top set of panels shows a MDI
continuum
intensity image while the map in the same position in the bottom set
shows a line-of-sight magnetogram of the sunspot in AR 9787.
The portion of the region shown here extends 219 Mm on each side.
For the purpose of this Figure (and the others in this
section), some spatial smoothing has been
applied to the travel-time maps.
}
\label{fig.hh_1}
\end{figure}

In step~5 we have used two general kinds of filters:
phase-speed filters and ridge filters. 
While not as commonly used as phase-speed
filters, ridge filters have
been used previously 
for $f$-mode studies
\citep[{e.g.}][]{Duvall2000, Gizon2002, Jackiewicz2007a, Jackiewicz2007b}
and recently for $p$ modes \citep{Jackiewicz2008, Braun2008}. 
The phase-speed filters used here are the same set of 11 filters 
(hereafter denoted ``TD1\,--\,TD11'') listed by \citet{Gizon2005} and
\citet{Couvidat2006} and commonly used
in time-distance analyses. The inner and outer radii
of the corresponding pupil quadrants are chosen so
that acoustic rays at a frequency of $\nu = $ 3.5 mHz
propagating from the focus to the edges of the pupil have
phase speeds (denoted by $w$) which span the
full width at half maximum (FWHM) of the squared filter.
The EW pupil quadrants for the 11 filters are shown in the bottom
set of panels in Figure~\ref{fig.hh_1}.
All of the phase speed filters used also remove the contribution of
the $f$ mode \citep{Braun2008}.

Figure~\ref{fig.hh_1} shows mean and EW travel-time shifts using
a frequency bandpass (step 6) between 2.5\,--\,5.5 mHz.
As is well known, there are distinct patterns of travel-time
shifts associated with the use of phase-speed filters.
In particular, the mean travel-time shift in the sunspot is positive
for the smallest values of phase speed (and mean pupil diameter;
{e.g.} TD1\,--\,TD3) and then switches sign for larger phase speeds 
(TD4 and beyond). The EW travel-time shifts also undergo a
similar change of sign. At the smallest (largest) phase speeds, the
EW travel-time differences are consistent with inflow-like (outflow-like)
perturbations centered on the sunspot. Remarkably, the switch in sign 
for both the
mean shift and the EW differences occurs between filters
TD3 and TD4.

\begin{figure}[]
\centerline{\scalebox{0.8}{\includegraphics{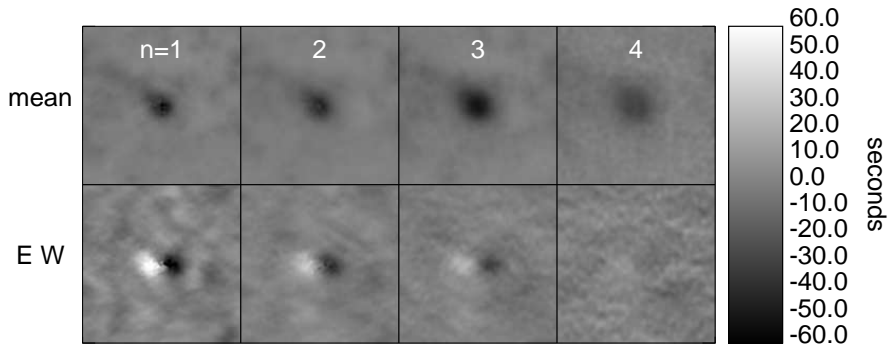}}}
\caption{
Maps of travel-time shifts $\delta \tau_{\rm{mean}}$ (top panels) and 
$\delta \tau_{\rm{EW}}$ (bottom panels) using ridge filters
over a frequency bandpass of 2.5\,--\,5.5 mHz for AR 9787 on
2002 January 24.
The columns 
indicate different radial orders as indicated. 
} 
\label{fig.hh_2}
\end{figure}

Figure~\ref{fig.hh_2} shows mean and EW travel-time shifts using
the same frequency bandpass (2.5\,--\,5.5 mHz) as shown
in Figure~\ref{fig.hh_1}, but
obtained with ridge filters isolating the $p_1$\,--\,$p_4$ ridges.
In contrast with the results obtained using phase-speed filters,
ridge filters show values of $\delta \tau_{\rm mean}$ which are
always negative within the sunspot, while $\delta \tau_{\rm EW}$ is consistent
with an outflow-like
perturbation. This is similar to results from time-distance
analyses as well as previous HH analyses
\citep{Braun2008}.

\begin{figure}[]
\centerline{\scalebox{0.8}{\includegraphics{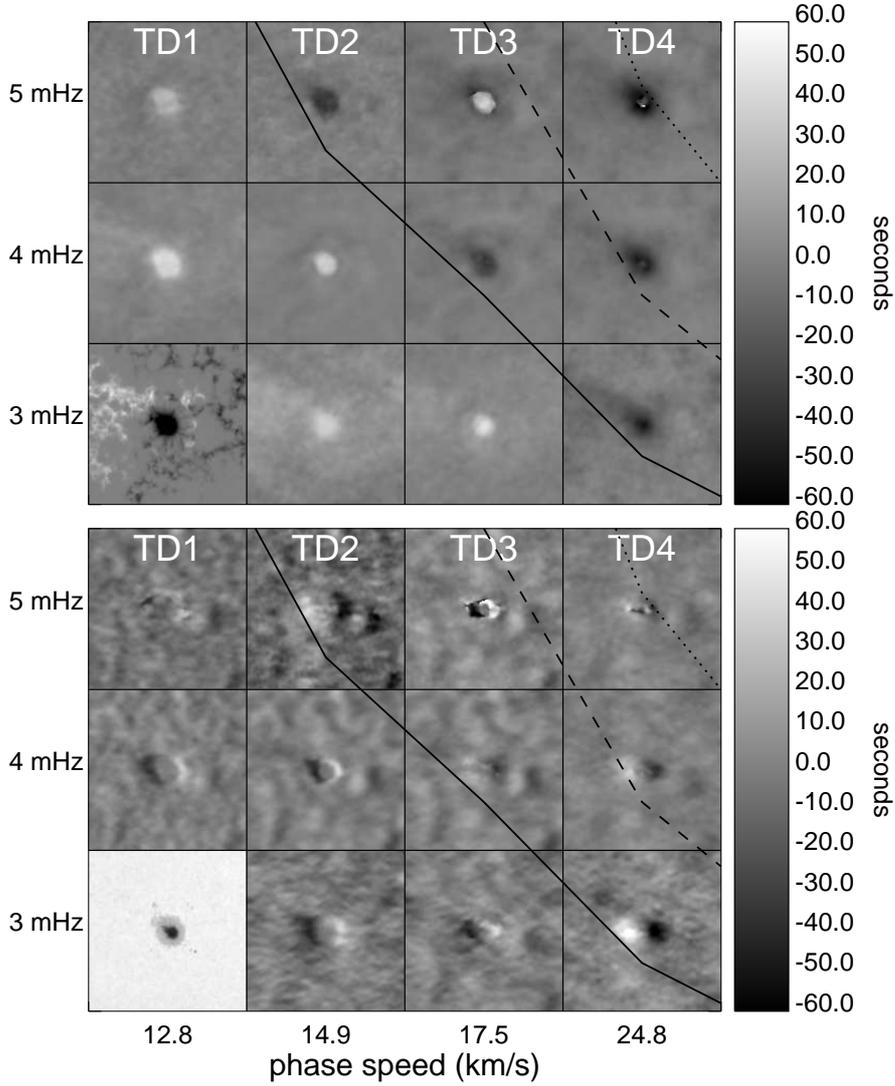}}}
\caption{
Maps of travel-time shifts $\delta \tau_{\rm{mean}}$ (top panels) and 
$\delta \tau_{\rm{EW}}$ (bottom panels) using phase speed filters and
1 mHz-wide frequency bandpasses for AR 9787 observed on 2002 January 24.
The columns of maps labeled TD1 through TD4 
indicate the phase-speed filter used,
while the rows indicate the frequency bandpass. The solid jagged line
running diagonally through the panels connects the location of
the $p_1$ ridge in the $\nu$-$w$  domain for each filter, with
the centers of the maps assigned to values of frequency and phase
speed as indicated
on the left and bottom edges of the plot. The dashed and dotted lines
indicate the locations of the $p_2$ and $p_3$ ridges respectively.
The map in the lowest-left position of the top set of panels shows a MDI
continuum intensity image while the map in the same position in the bottom set
shows a line-of-sight magnetogram.
}
\label{fig.hh_3}
\end{figure}

Motivated by recent studies which show frequency variations
of travel-time shifts observed in active regions
\citep{Braun2006, Couvidat2007, Braun2008}, we also employed
narrow frequency bandpasses to both the phase-speed and ridge
filters.  Figure~\ref{fig.hh_3} shows mean and EW travel-time shifts using
phase-speed filters TD1\,--\,TD4 in conjunction with 1-mHz wide
frequency filters centered at 2, 3, 4, and 5~mHz.
Positive mean travel-time shifts, and EW travel-time differences
consistent with inflows,
are observed primarily in frequency
bandwidths that are centered below the $p_1$
ridge, shown by the solid line in Figures~\ref{fig.hh_3}.
There is one instance (filter TD3 at 5mHz, which lies
immediately below the $p_2$ ridge) which also produces
a positive mean shift and inflow-like signature. All
other filters (including TD5\,--\,TD11 not shown) show
negative mean travel-time shifts and outflow-like 
signatures. 

\citet{Braun2008} find that a condition for producing positive travel-time shifts such as in Figure~\ref{fig.hh_3} appears to be
a disproportionate contribution to the correlations of 
wave power from the low-frequency wing of the $p_1$ ridge 
relative to the high-frequency wing.
We note that recent work by \citet{Moradi2008} is relevant to these issues.

\subsection{Ridge and Off-Ridge Filtering}

\begin{figure}
\begin{center}
\includegraphics[width=28pc]{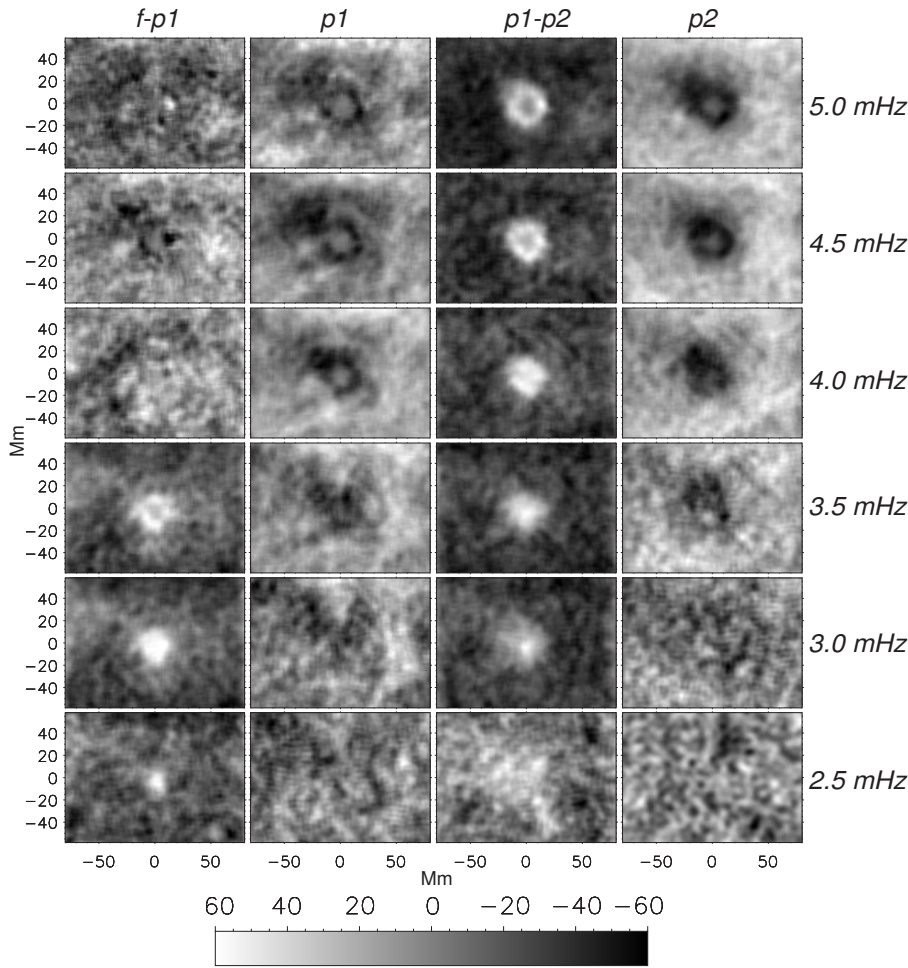}
\caption{Travel-time perturbations $\delta \tau_{\rm{mean}}$ for isolated sunspot AR~9787 obtained for various filtering schemes. The colorbar is in units of seconds. The skip distance is equal to 11.64 Mm. From the left, the columns respectively have filters applied as follows: pass-filter centred between the  $f$ and $p_1$ ridges; pass-filter centred on the $p_1$ ridge; 
pass-filter centred between the $p_1$ and $p_2$ ridges; pass-filter centred
on the $p_2$ ridge. From the bottom, the rows have  
bandpass filters centred on 
$2.5$, $3.0$, $3.5$, $4.0$, $4.5$, and $5.0$~mHz  
respectively.  More details are given
in the text.}
\label{fig:noaa_filters11.664}
\end{center}
\end{figure}

Following \cite{Braun2006}, \citet{Braun2008} and \citet{Thompson2008}, we 
study the sensitivity to filtering of mean travel-time perturbations
measured in the vicinity of isolated sunspot AR~9787 relative to 
the surrounding quiet Sun, using a centre-to-annulus geometry and a 
skip distance of $11.7$~Mm (Figure~\ref{fig:noaa_filters11.664}).
For each row a bandpass filter was used to 
select data within a 1~mHz frequency band with a $0.1$~mHz Gaussian roll-off, 
centred (from bottom to top) at $2.5$, $3.0$, $3.5$, $4.0$, 
$4.5$ and $5.0$~mHz.
These were combined for each column (left to right) with filters selecting 
the data from in between $f$ and $p_1$ ridges, from the $p_1$ ridge, in between the $p_1$ and $p_2$ ridges, and the $p_2$ ridge. 
The filters were constructed as follows: 
at constant frequency we apply a filter that takes the value of unity at the
horizontal wavenumber corresponding to either a particular ridge, 
{e.g.} $p_1$ (a ``ridge filter''), or a mid-point between the 
adjacent ridges, {e.g.} $p_1-p_2$ (an ``off-ridge filter''). 
On either side of this centre line the filter has a Gaussian roll-off with 
half width at half maximum (HWHM) equal to $0.32$ times the distance to the 
neighbouring ridge on that side for the ridge filter and with HWHM equal
to $0.63$ times the distance to the adjacent ridge in the case
of the off-ridge filter. 
No phase-speed filter was applied. 

The cross-correlation functions were estimated using center-to-annulus geometry with annuli taken to be one data pixel wide. Travel-time perturbations were measured using the definition of \citet{Gizon2004b}.

In agreement with \citet{Braun2006} and \citet{Braun2008}, as illustrated in 
Figure~\ref{fig:noaa_filters11.664}, we observe a positive travel-time
perturbation in the region beneath the $p_1$ ridge, but we also find such 
a signal between the $p_1$ and $p_2$ ridges, and find that on the 
$p_1$ ridge the positive perturbation is absent. 
Our tentative conclusion is that the positive travel-time perturbation
signal arises only in the regions between the $p$-mode ridges; and that
the travel-time perturbations associated with the data on the ridges 
themselves are all consistently negative. 
Similar, though noisier, results were obtained for the Gabor-wavelet fitting
travel-time definition. The data appear to suggest that measured travel-time perturbations are very sensitive to
the part of the wave-propagation diagram selected during the filtering stage.

\section{Moat Flow Inversion (Ridge Filters)}
\label{jack}

Here we invert travel times to obtain the flows around the sunspot in AR~9787. We restrict ourselves to ridge filtering and TD travel times.  We then compare the inferred flows with the velocities of the moving magnetic features (MMFs) in the moat (Section~\ref{ss:thedata}).

 The details of the travel-time inversion can be found in \citet{Jackiewicz2008}.  In summary, we measure center-to-quadrant travel-time differences using the method of  \citet{Gizon2004b}, after correcting for reduced power in magnetic regions.  Using the same definition of travel times, Born sensitivity kernels are computed \citep{Birch2007a}. We input the travel times and the sensitivity functions, as well as the covariance of the travel times, into a three-dimensional subtractive optimally localized averages (SOLA) inversion procedure to infer the vector flows at several depths. The procedure also provides good estimations of the resolution and the noise levels, which are  important for any interpretation. We note that neither the modeling nor the inversion takes into account the magnetic field.

\begin{figure}
  
 \centerline{  
    \begin{minipage}[T]{0.48\textwidth}
      {\includegraphics[width=0.96\textwidth]{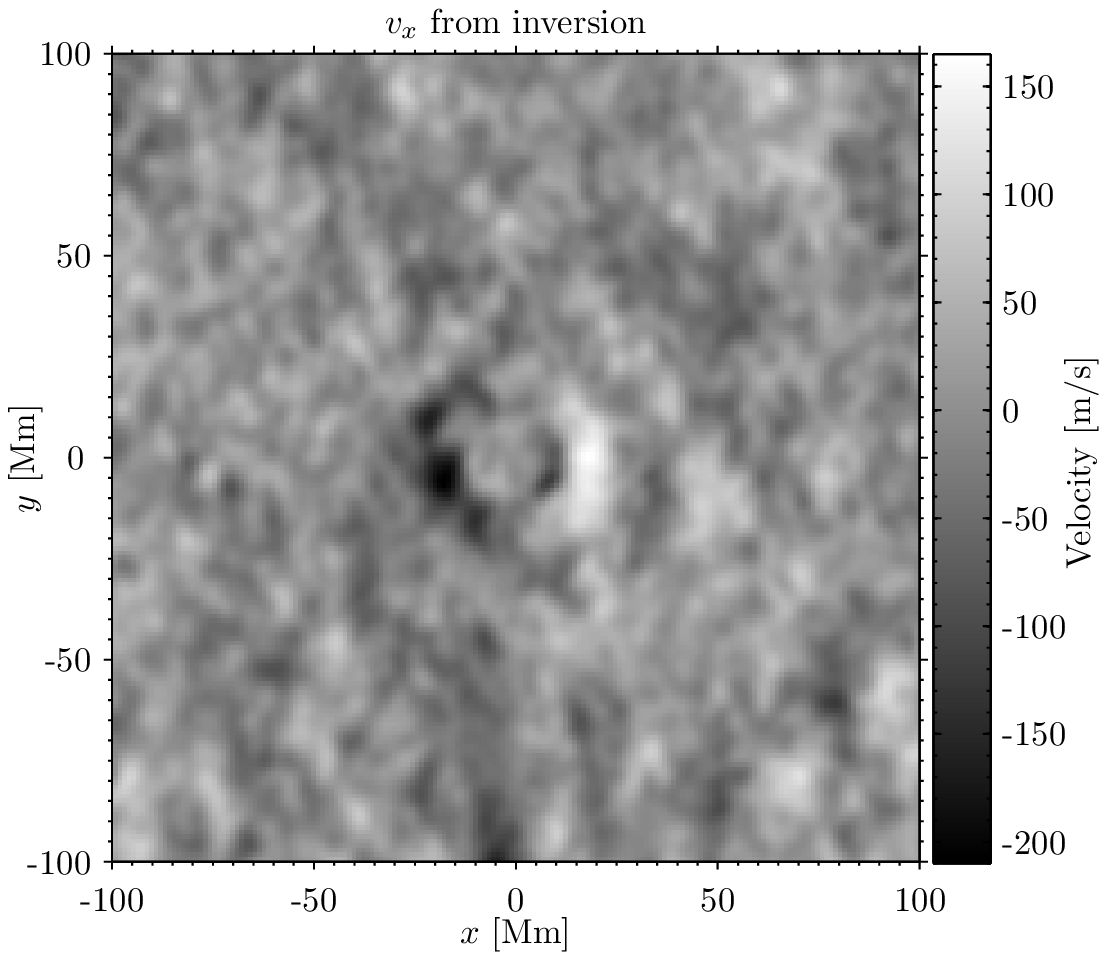}}
  \end{minipage}
  \begin{minipage}[T]{0.48\textwidth}
    {\includegraphics[width=0.96\textwidth]{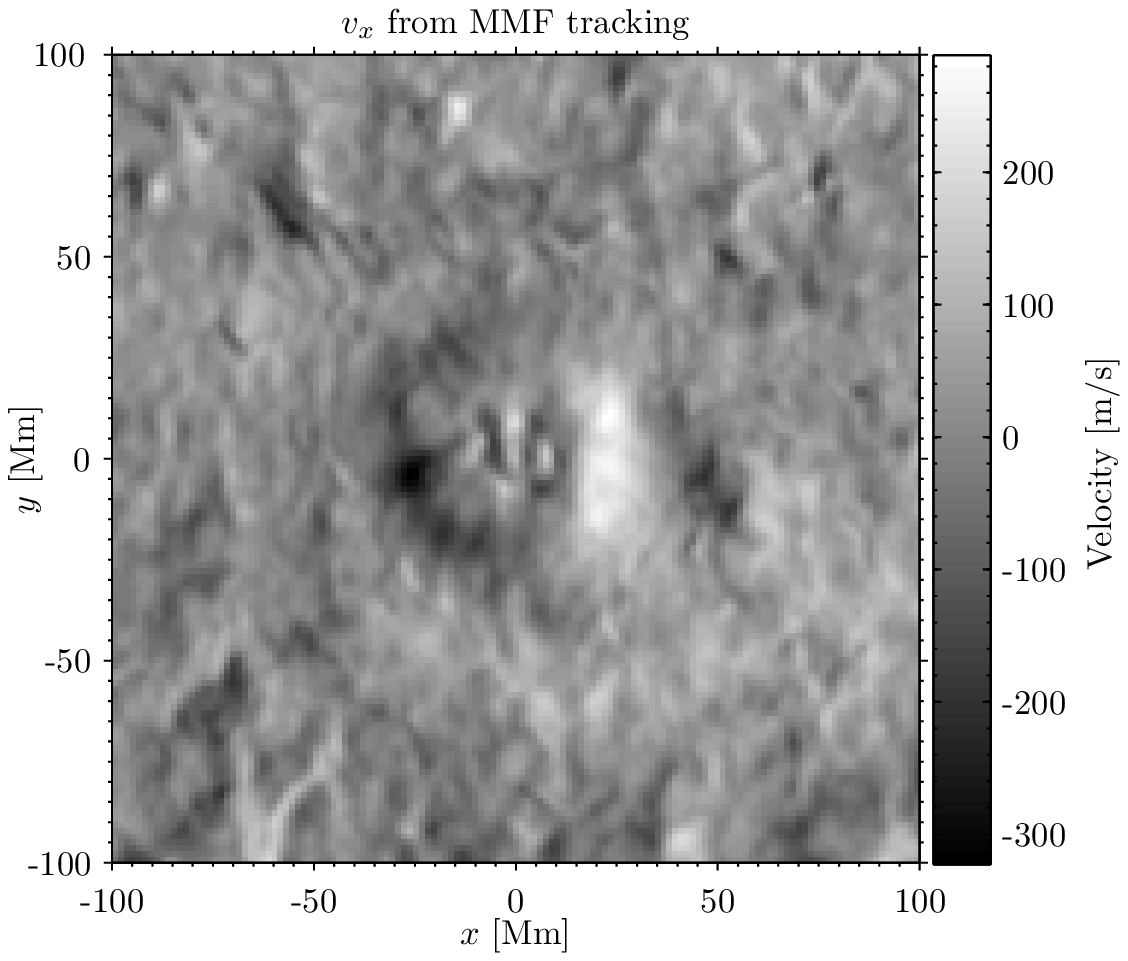}}
  \end{minipage}}
\vspace{0.5cm}
 \centerline{  
    \begin{minipage}[t]{0.48\textwidth}
      {\includegraphics[width=0.96\textwidth]{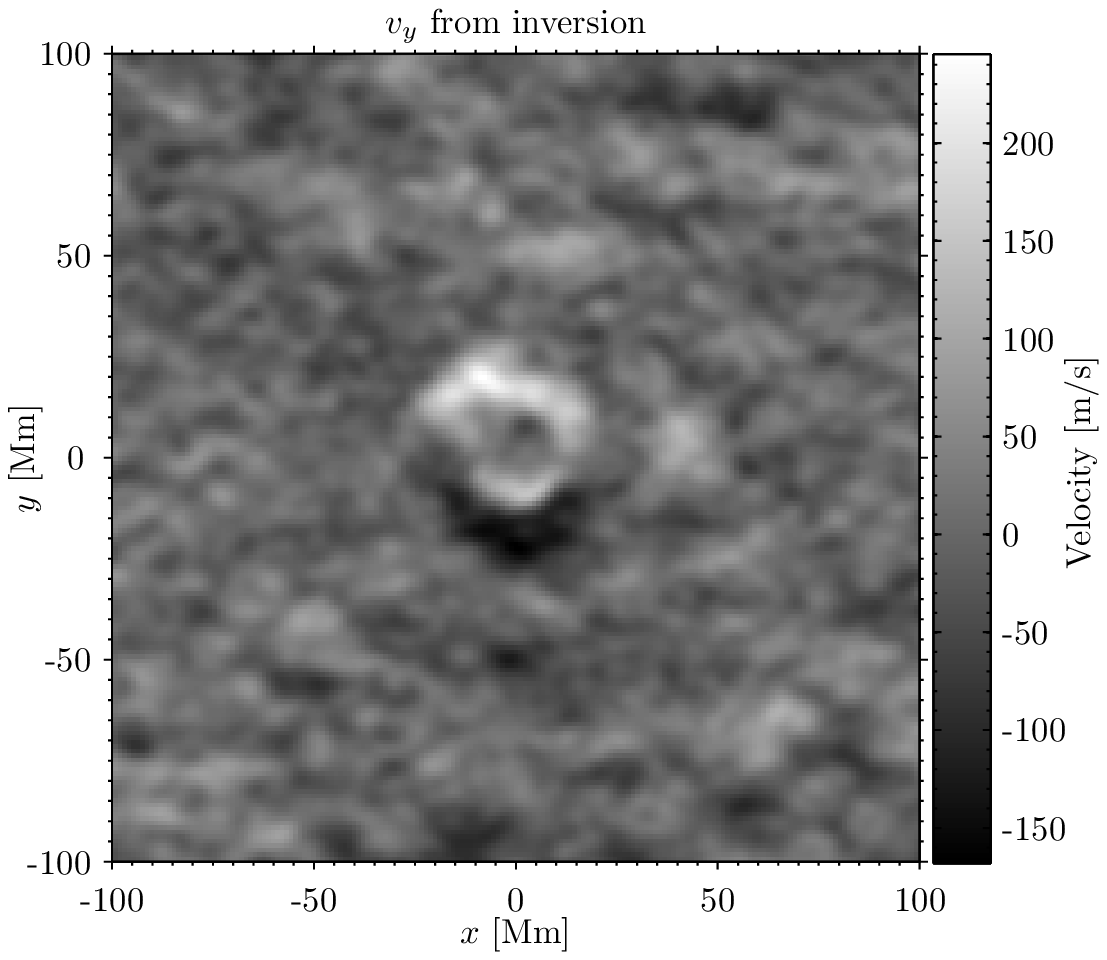}}
  \end{minipage}
  \begin{minipage}[t]{0.48\textwidth}
    {\includegraphics[width=0.96\textwidth]{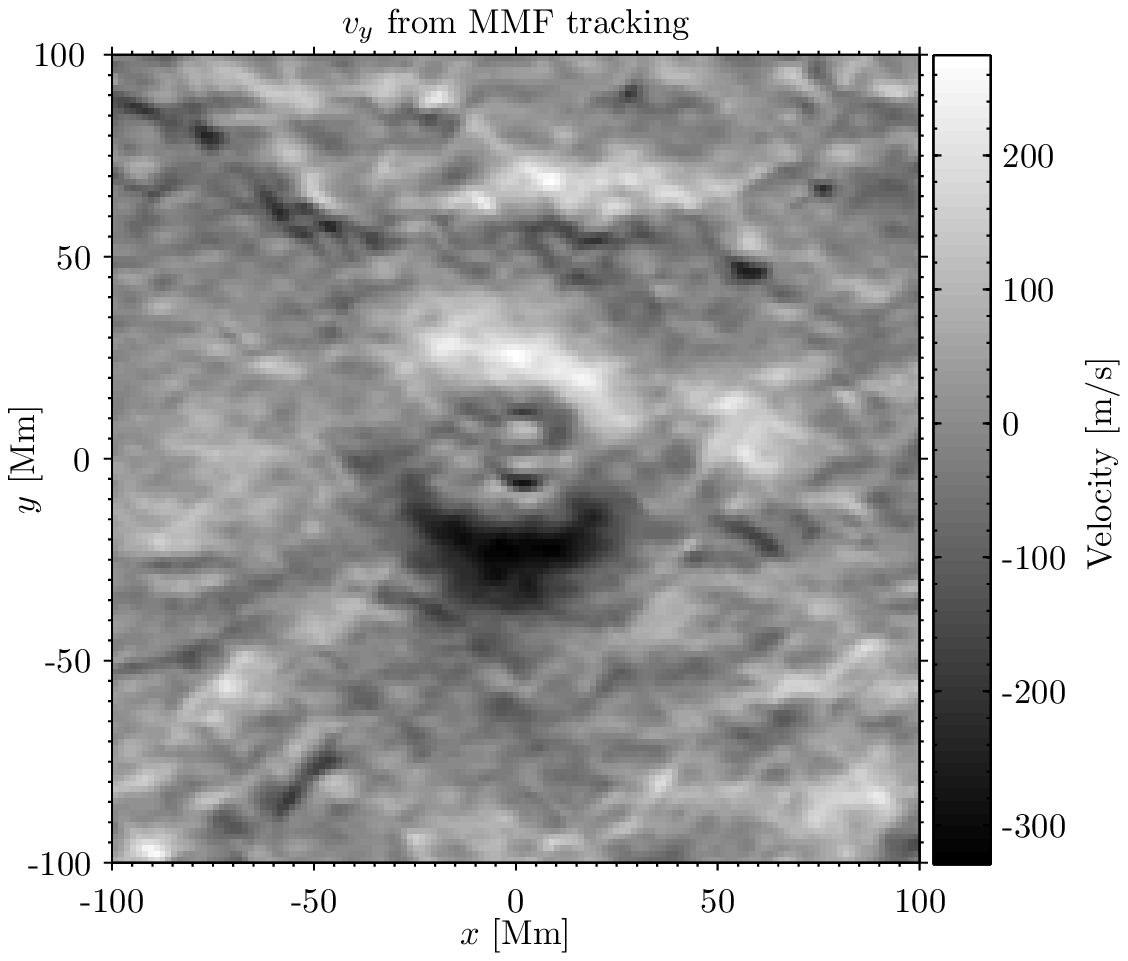}}
  \end{minipage}}
\vspace{0.5cm}
  \centerline{  
    \begin{minipage}[T]{0.48\textwidth}
      {\includegraphics[width=0.96\textwidth]{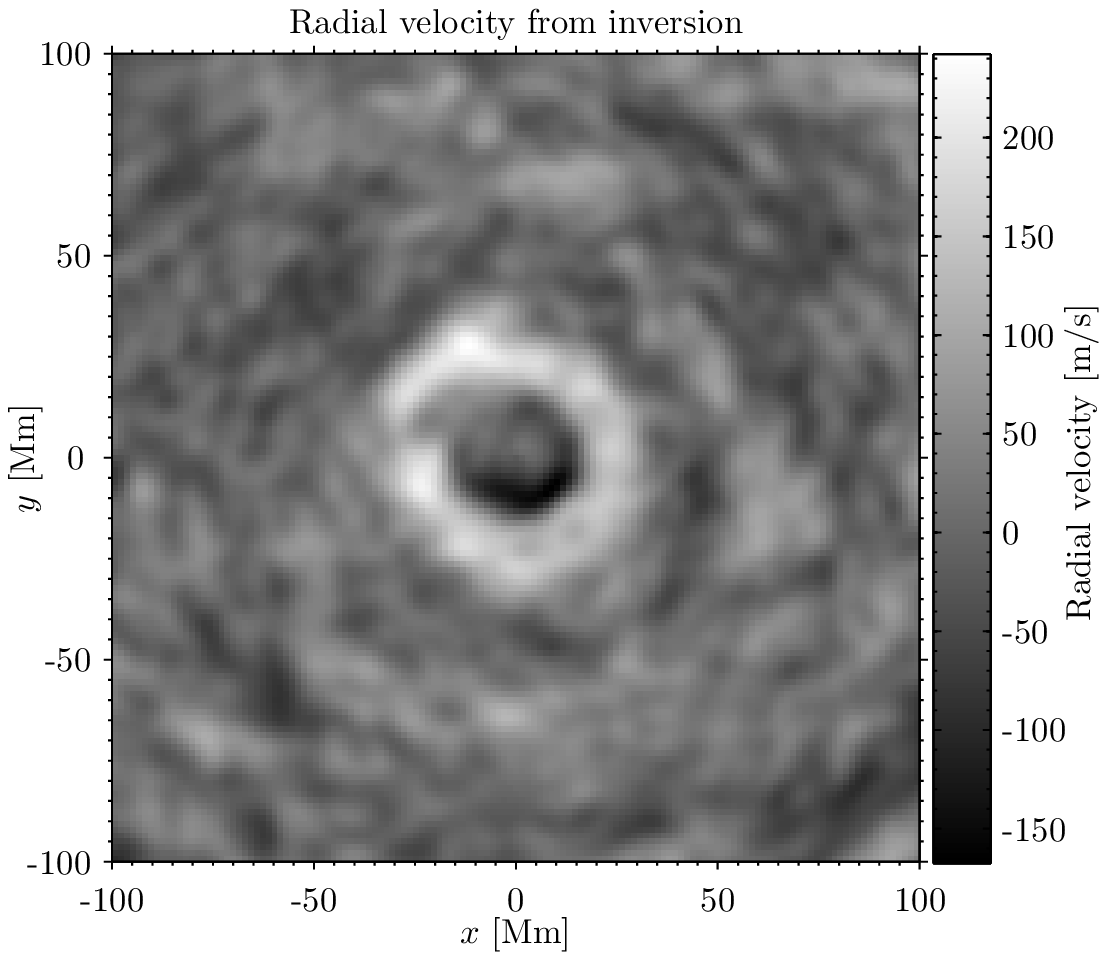}}
  \end{minipage}
  \begin{minipage}[T]{0.48\textwidth}
    {\includegraphics[width=0.96\textwidth]{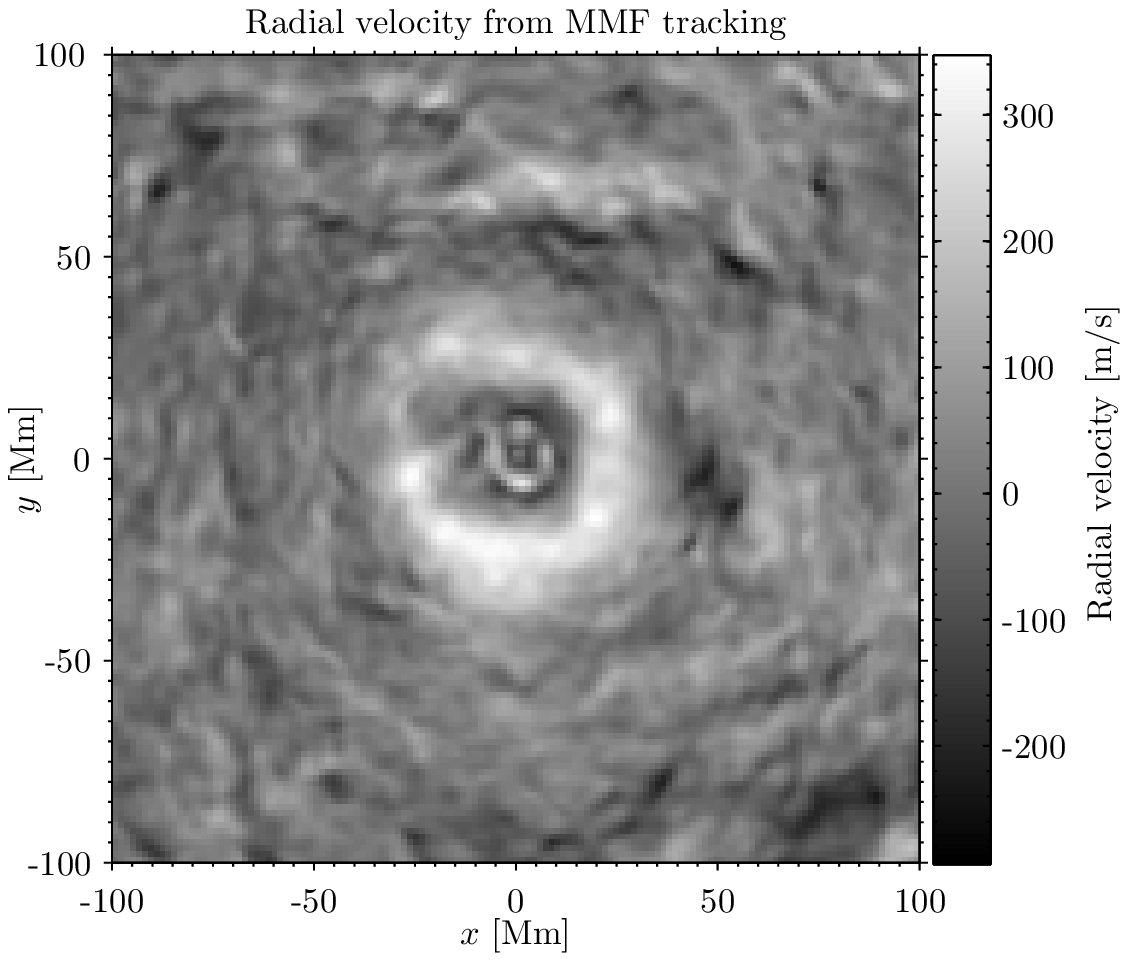}}
  \end{minipage}}
  \caption{Comparison of near-surface flows around the sunspot from time-distance inversions and MMF tracking. The left column shows flows obtained from inversions at a depth of 1~Mm beneath the surface and averaged over 7 days. The right column are the flows obtained from MMF tracking averaged over the same 7 days. Each set of flows was obtained with approximately the same resolution/smoothing per pixel. The top row is the $x$ component of the velocity, and the middle row shows the $y$ component of the velocity. The bottom row compares the radial velocity (from the center of the sunpot) of the two measurements. An outward moat flow is seen for each case.}
  \label{fig:moat_matrix}
\end{figure}

We have inverted the middle seven days of data from the nine day set. We obtain flow maps for several depths, extending down to about 5~Mm beneath the surface. In Figure~\ref{fig:moat_matrix} we compare the inversion results near the surface (left column) to the motion of the moving magnetic features (MMFs, right column). For both sets of maps, the flows are averaged over 7 days. Furthermore, they are  approximately of the same horizontal resolution ($\sim 6$~Mm). For this particular figure, we choose to study the inferred time-distance flows taken at a depth as near to the surface as we can achieve, about $1$~Mm below.

The bottom row of Figure~\ref{fig:moat_matrix} compares the radial velocities 
derived from TD helioseismology and MMF tracking. We see for each case quite clearly a strong outflow extending beyond the penumbra (20~Mm) of several hundred m/s, known as the moat flow. The overall features of the flows from  both methods are quite similar, even the slight `knob' on the northeast quadrant of the sunspot moat. A look in the quiet Sun reveals other similarities. The correlation coefficient between the two maps is about $0.65$. The magnitude of the TD surface flows is about 20\% less than that revealed by MMF tracking.  This can be due to many factors, such as the implied depth at which we are comparing not being equal, inaccurate travel-time sensitivity kernels, or the magnetic field affecting the inversion results through the travel times, among others. The estimated noise in the time-distance maps is about 5~m/s.
These results are consistent with a previous $f$-mode TD study of sunspot moat flows \citep{Gizon2000}.

\begin{figure}
\vspace{0.2cm}
  \centerline{\includegraphics[width=.95\textwidth]{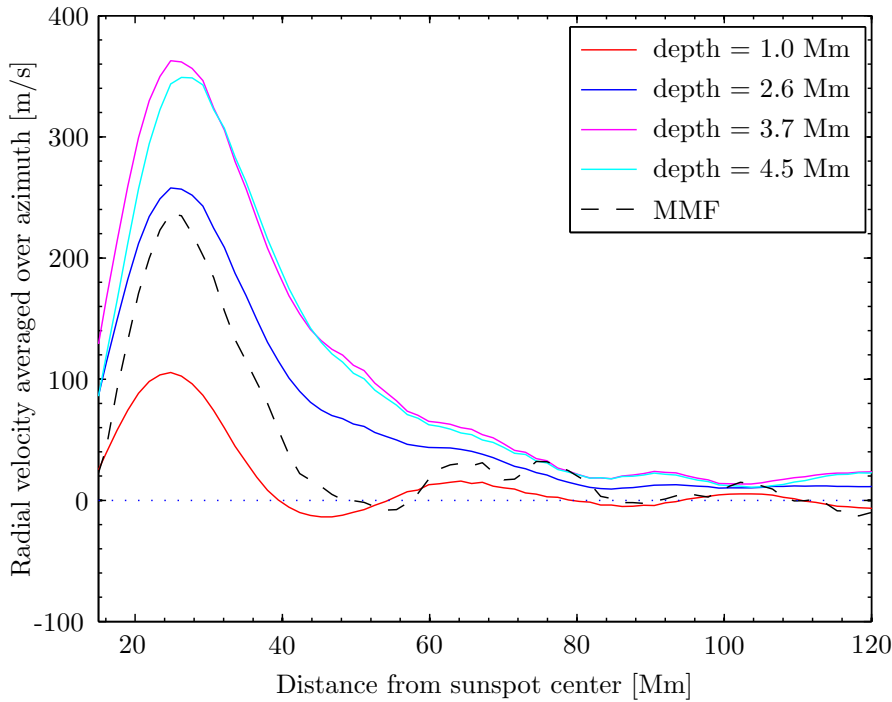}}
  \caption{Azimuthally-averaged radial flows from the sunspot center at different depths over 7 days, obtained from a time-distance inversion.  The radial velocity from the MMF tracking is shown by the dashed line. In this plot only the region from the edge of the penumbra outward is shown. The results within the sunspot (distance less than $20$ Mm) cannot be reliably interpreted.}
  \label{fig:v_azim}
\end{figure}

Since we do not know the exact depth where the small magnetic features are rooted, it is useful to compare their motion with inverted flows at several depths. Since the sunspot is very nearly circular, we may first average the radial flows azimuthally about the sunspot center. In Figure~\ref{fig:v_azim} we plot the azimuthally-averaged radial flows for several depths versus the distance from the center of the sunspot. Also shown for comparison is the averaged MMF velocity. The moat flow, which extends to about 45~Mm, is seen at all depths. The inferred flows get stronger with depth and begins to level off in strength at about $4.5$~Mm below the surface. The MMF velocity is consistent with these flows at depths of between 1 and $2.6$~Mm beneath the surface.  The shape of the moat flow is similar for all cases.

\section{Sound Speed versus Wave Speed}

The sound speed, $c$,  is the  speed  of sound. In this paper we make an important distinction between $c$ and the speed at which waves are inferred to have travelled under the assumptions  of the helioseismic inversions.  Different inversions will include different physics.
A common example is the assumption that a travel-time perturbation associated with a sunspot can be modelled purely by an equivalent small-amplitude sound-speed perturbation. The sound-speed inferred under such an assumption is at best a local wave speed, and need not even reflect the sign of the sound speed perturbation. For this reason, we use the notation $c_{\rm w}$ to denote the inferred wave speed.  

 \citet{Lin2008} studied the exact meaning of $c_{\rm w}$ in the case of ring-diagram analysis. They find that their inversions return perturbations in $c^2_{\rm w} = \Gamma_1 P_{\rm tot}/\rho$, where $P_{\rm tot}=P+P_{\rm mag}$ is the sum of the gas and magnetic pressures. The inferred wave speed $c_{\rm w}$ has two components: the sound speed, $c$, and a magnetic component.

What we have said about sound speed versus wave speed, also applies to any chosen physical quantity versus inferred quantity. For example, the next section presents an inversion for wave slowness, $s_{\rm w}=1/c_{\rm w}$, using traveltime sensitivity kernels for inverse sound speed $s=1/c$.


\section{Wave-Speed Inversion (Phase-Speed Filters)}\label{zharkov}

The data was reduced using standard time-distance helioseismology techniques. First, the data was preprocessed by applying an amplitude modulation correction as described by \citet{Rajaguru2006}, followed by applying a high-pass filter at $1.7$ mHz in order to remove the supergranulation and a low-pass filter at $5.1$ mHz to remove signal above the acoustic cut-off frequency. We then apply a Gaussian phase speed filter to select waves with horizontal phase speed $w_i$ near the value corresponding to skip distance $\Delta_i$ given by ray theory. We define 12 such filters $F_i$ corresponding to different distances $\Delta_i$.  These phase-speed filters are similar to the set of filters used in Section~\ref{sec.psfilt} and commonly used
in time-distance analyses. Each filter is applied by pointwise multiplication of the Fourier transform $\psi(k_x, k_y, \omega)$ of the observed velocity data.

We use a centre-to-annulus geometry to compute the cross-covariance $C({\bf r}, \Delta_i, t)$, where ${\bf r}$ is the center of the annulus and $t$ is the time lag. The annulus width is $4.5$~Mm. A reference cross-covariance function, $C^{\rm ref}$, is obtained by spatial averaging $C$ over a quiet Sun area. Wave travel times are then extracted by fitting a Gabor wavelet to the positive- and negative-time branches of $C$ \citep{Kosovichev1997}. The wavelet has five parameters:  the central frequency, the width and amplitude of the envelope, and the group and phase travel times. We denote by $\tau_+$ and $\tau_-$ the measured phase travel times for the positive- and negative-time branches of $C$ respectively. The reference travel times for the quiet Sun are similarly defined using $C^{\rm ref}$. The phase travel time perturbations, $\delta\tau_+$ and $\delta\tau_-$, are defined as the difference between the measured and reference travel times. As we are interested in wave-speed perturbations only, we consider mean travel-time perturbations, $\delta \tau_{\rm mean} = (\delta\tau_+ + \delta\tau_-)/2$.

For the forward problem we use sensitivity kernels estimated using the first-order Rytov approximation \citep{Jensen2003}. These kernels, $K^s$, relate mean travel-time perturbations, $\delta\tau_{\rm mean}$, to inverse sound speed perturbations, $\delta s=\delta(1/c)$, of a quiet-Sun model. In the sunspot region, we have
\begin{eqnarray}
\delta \tau_{\rm mean} ( {\bf r}, \Delta_i) =  \int_S {\rm d}^2{\bf r'} 
\int_{-d}^0 {\rm d}z \; K^{s}({\bf r- r}', z;\Delta_i)\; \delta s_{\rm w} ({\bf r}', z),
\label{eq_TT_Kernel}
\end{eqnarray}
where $S$ is the area of the region, $d$ is its depth. The quantity $\delta s_{\rm w} = s_{\rm w} - s$ is the equivalent  change in the local wave slowness caused by the sunspot.

We invert for $N=14$ layers in depth located at $[z_1, \cdots z_{N}] =  [ 0.36$, $1.2$, $2.1$,				 	$3.3$, $4.7$, $6.4$,
					$8.6$, $11.2$, $14.3$,
					$17.8$, $21.8$, $26.3$,
					$31.4$, $37.0]$~Mm.
We use a multi-channel deconvolution algorithm \citep{jensen98, Jensen2001} enhanced by the addition of horizontal regularization \citep{Couvidat2006}. 
The above equation is Fourier transformed with respect to two-dimensional position ${\bf r}$.
For each wavevector ${\bf k}$, we  define 
$d_i   =  \delta \tau ({\it \vec{k}}, \Delta_i )$, 
$G_{ij}  =K({\it \vec{k}}, z_j; \Delta_i )$, 
and $m_j  =\delta s({\it \vec{k}}, z_j)$,
and the corresponding vector ${\bf d}$, matrix $G$, and vector ${\bf m}$.
Then for each ${\it \vec{k}}$ we solve for the vector ${\it \vec{m}}$ that minimizes
\begin{eqnarray}
 \Vert (\vec{d}-G \vec{m})  \Vert^2 +\epsilon \Vert L\vec{m} \Vert^2  ,
\end{eqnarray}
where $L$ is a regularization operator and $\epsilon({\bf k})$ is a positive regularization parameter. In this work we apply more regularization at larger depths, to which travel times are less sensitive, by setting
$L={\rm diag}(c_{1}, c_2, \ldots c_{N})$, where $c_j = c(z_j)$ is the sound speed in the $j$-th layer of the reference model. 
We regularise small horizontal scales by taking $\epsilon(\vec{k})= 2\times 10^3 (1+|\vec{k}|^2)^{100}$.	
					
\begin{figure}
\begin{center}
\includegraphics[width=25pc]{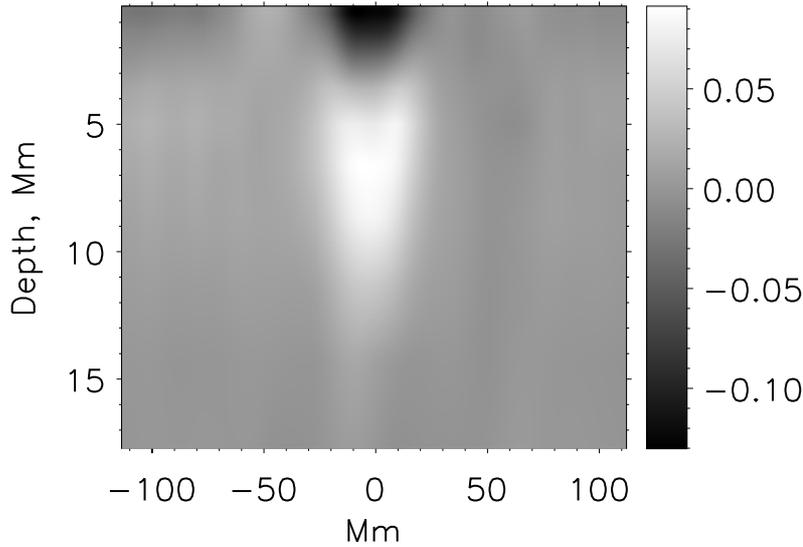}  
\caption{Relative wave speed perturbation, $\delta c_{\rm w} /c$,  obtained for AR 9787 using time-distance helioseismology and phase speed filtering. }
\label{fig:avessInversion}
\end{center}
\end{figure}

Figure~\ref{fig:avessInversion} shows the result of the inversion, expressed in terms of the relative wave-speed perturbation $\delta c_{\rm w}/ c$. We see a two-layer structure: a region of decreased wave speed (down by $-13\%$) situated directly underneath the surface and a region of increased wave speed (up to $9\%$) starting from a depth of approximately $3$~Mm.
This is consistent with other time-distance inversions of travel times using phase-speed filters, {e.g.}, those of \citet{Kosovichev2000}.

\section{Ring-Diagram Analysis}\label{bogart}
\begin{figure}
\hspace{-0.5cm}
\includegraphics[width=12cm]{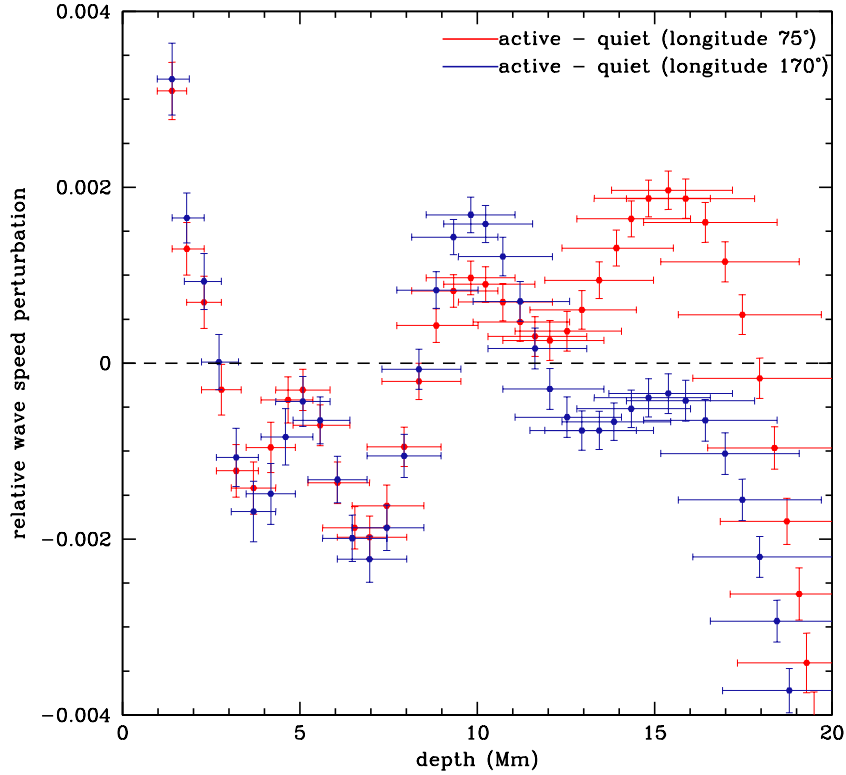}
\caption{SOLA inversions for the depth dependence of the relative difference in wave speed, $\delta c_{\rm w}/c$, between the region around AR 9787 and each of two comparison quiet regions at the same latitude but different longitudes. The inversions are based on fits to power spectra for $5.7$ days of tracked data for each region. The differences are in the sense $\delta c_{\rm w} = c_{\rm w, active} - c_{\rm w, quiet}$. The active region is seen to have a positive wave-speed anomaly very near the surface relative to the quiet regions.}
\label{rbcb-1}
\end{figure}

We have used ring diagrams to analyze the mean structure of the
region containing AR~9787 compared with quiet-Sun structure. To do so, we
use the techniques described in \citet{Basu2004}. We invert the
differences in the ring-diagram fit parameters between the spectra of the
active region and those of suitable selected quiet-Sun regions. In this
case, two quiet-Sun regions were chosen at the same latitude as that
of the active region ($-7^\circ$) and at Carrington longitudes $170^\circ$ and $75^\circ$
(the active region is at longitude $130^\circ$). Each region is independently
tracked in a time interval of $5.7$~days  centered on its
central meridian crossing, so any geometrical differences in the spectra
due to foreshortening or geometric image distortion are very nearly
canceled out. The regions chosen for analysis are $15^\circ$ in diameter,
so the results apply to a spatial mean over these areas (with an
unknown weighting function). The average results from the two comparison regions is shown in  Figure \ref{rbcb-1}. It is clear that
there is a region of negative wave-speed anomaly under the active region
between $3$~Mm and $8$~Mm in depth, with a turnover to positive wave-speed
anomalies both above and below this region, and yet another turnover to
negative anomalies at depths greater than about $17$~Mm. This behaviour is typical of that seen for other active regions \citep{Bogart2008}, although the changes at the surface and deeper than $17$~Mm are unusually pronounced in this case.

\begin{figure}
\includegraphics[width=12cm]{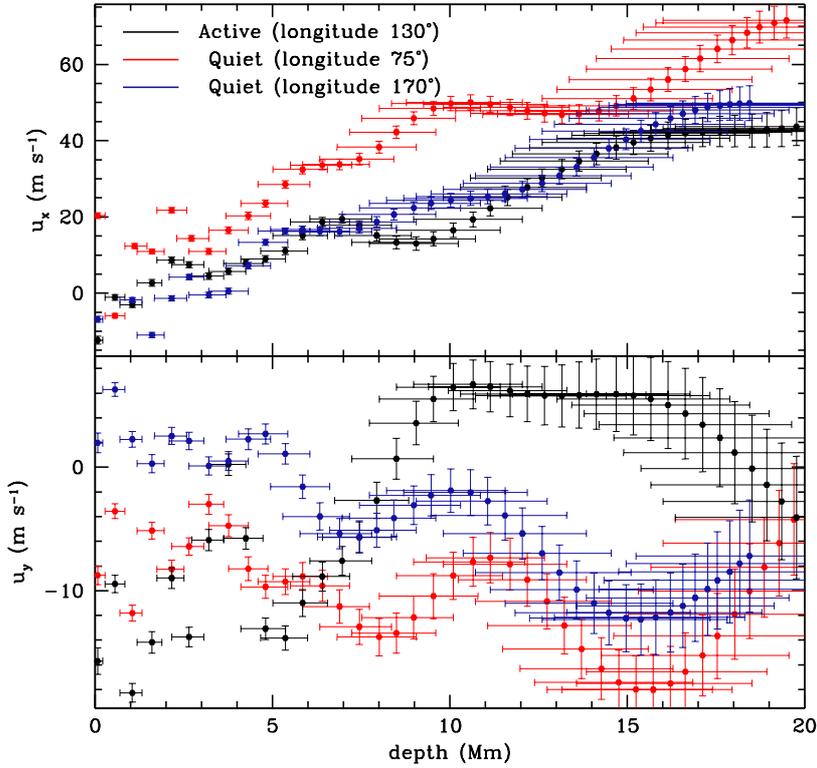}
\caption{SOLA inversions for the depth dependence of the mean zonal
($u_x$) and meridional ($u_y$) flows over each of the three regions analyzed
with ring-diagram analysis. The zonal rates are relative to the tracking
velocity, which was 54 m/s less than the Carrington rate at this latitude.}
\label{rbcb-2}
\end{figure}

We can also infer the mean flow speeds at depth for the active regions
and the comparison quiet regions directly from the fitted parameters to
the ring-diagram spectra. These are shown in Figure \ref{rbcb-2}. There is no
evident anomalous zonal flow through the active region; indeed, the zonal
flow structure is remarkably similar to that of the preceding comparison
region at longitude $170^\circ$. There does appear to be an anomaly in the
structure of the meridional flow, however, with a substantial shear
at depths greater than $7$~Mm, the flow being poleward near the surface (the
region is in the southern hemisphere) and equatorward at greater depths.
It is especially marked if the mean meridional velocity at the active
region's latitude is negative at depth, as the two comparison regions
suggest, but this needs to be verified by averaging over more longitudes.

\section{Numerical Forward Modeling}
In various circumstances it has been shown that the cross-covariance
is closely related to the Green's function.  This allows us to 
characterize the interaction of arbitrary wavepackets with the sunspot 
from the MDI observations.
The sunspot discussed in this paper, being observed over nine days and almost
axisymmetric, is ideally suited to such a study. In preliminary work,
\citet{Cameron2008} considered the cross-covariance between
the Doppler signal averaged along a great circle 40~Mm from the
centre of the sunspot and the Doppler signal at each point in a region
 surrounding the sunspot. The data had been $f$-mode ridge filtered.

A numerical simulation was then performed of the propagation an $f$-mode
plane wave packet beginning 40~Mm from a model sunspot. 
The background atmosphere is Model~S of \citet{jcd96}, 
stabilized with respect to convection. The sunspot model used was a
simple self-similar model in the vein of \citet{Schlueter1958}.
The half-width of the vertical magnetic field at the surface was taken to be
10~Mm (as in AR 9787) and field strengths of 2000, 2500 and 3000~G were considered.
The SLiM code \citep{Cameron2007} was used to perform
the simulations.

\begin{figure}
\begin{center}
\includegraphics[width=10cm,clip]{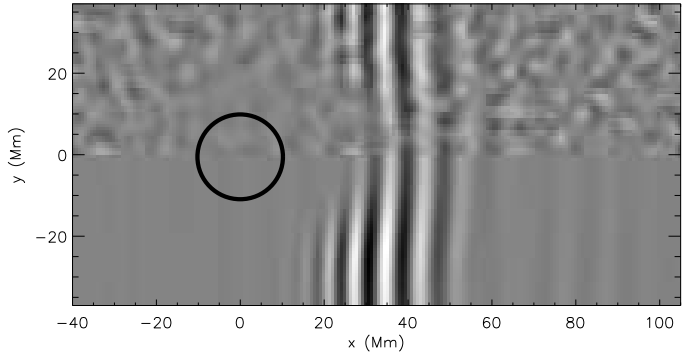}
\end{center}
\caption{
Comparison between the observed cross-covariance and a numerical simulation. 
The circle indicates the location of the sunspot in AR~9787.
(Top half) $f$-mode cross-covariance between the MDI Doppler signal averaged over a 
line at $x=-40$~Mm and the Doppler signal at each point. The correlation 
time-lag is 130~min, large enough for wave packets to traverse the sunspot.
The cross-covariance is averaged over 9 days and uses the assumed azimuthal symmetry of the sunspot to reduce noise. 
(Bottom half) SLiM numerical simulation of an $f$-mode wave packet propagating in the 
+$x$ direction through a model of AR~9787 with a peak magnetic field of 3~kG.
(From \citet{Cameron2008})
}
\label{fig.slim}
\end{figure}

\begin{figure}
\begin{center}
\includegraphics[width=10cm]{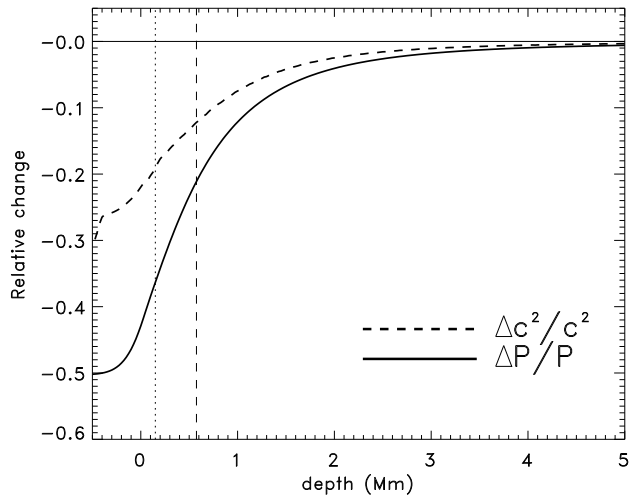}
\end{center}
\caption{
Structure of the 3~kG sunspot model used in the simulation of Figure~\ref{fig.slim}. The thick solid curve shows the relative change in gas pressure, $P$, with respect to the quiet Sun, measured along the sunspot axis.  The thick dashed curve shows the relative change in the squared sound speed, $c^2$, which is also an estimate of the relative change in temperature. The vertical dashed line indicates the depth at which the sound speed and the Alfv\'en velocity are equal. The vertical dotted line indicates the depth  at which the sound speed is equal to its zero-depth quiet-Sun value.
}
\label{fig.slim_model}
\end{figure}

The top half of Figure~\ref{fig.slim} shows the observed cross-correlation with time-lag $130$~min.  The bottom half of the figure shows the results of
the simulation at time $130$~min. The match between the observations and
simulation is quite good, in this case for a sunspot with a peak field strength of
3000~G. The match was not as good for peak field strengths of 2000~G and 2500~G. This then places a helioseismic constraint on the magnetic field of the
spot. Whereas this constraint makes sense, it cannot be assessed directly 
using MDI magnetograms, which are not reliable in sunspot umbrae. 
The partial absorption of the waves (reduced cross-covariance amplitude) was explained in terms of 
mode conversion into slow magneto-acoustic waves that propagate down the 
sunspot, as predicted by, {e.g.},  \citet{Cally2000}.
Full details of this work are given in \citet{Cameron2008}.

For the 3 kG sunspot model, Figure~\ref{fig.slim_model} shows
 the relative change in the gas pressure 
and the square of the sound speed along the sunspot axis with respect to the quiet Sun model.
Both perturbations become very small (less than 1\%) deeper than a depth of about 4~Mm. 
The temperature, closely related to the squared sound speed,  is reduced at all
depths within the sunspot.
At the surface the relative decrease in temperature is around 18\%. 
The reduction in gas pressure is larger with a 36\% decrease. 
The vertical dashed line
indicates that the sound  speed is equal to the Alfv\'en speed ($c=a$)
at a depth of approximately 580~km. The
$c=a$ level is where mode conversion is expected to occur. We also plot
the depth at which the sound speed is equal to the surface
quiet-Sun sound speed. This gives a rough indication of the Wilson
depression, in this case a rather low 170~km.

\section{Discussion: The Elusive Structure of Sunspots}

\subsection{Problematic Travel Times}
\label{flowbug}

Here we use Born approximation based forward modeling to test if the flows estimated from inversions of the travel-times obtained using ridge filters (Section~\ref{jack}) are consistent with the travel-times measured using phase-speed filters (Section~\ref{doug}).

To carry out the forward modeling we employ sensitivity functions, kernels, computed using the method of \citet{Birch2007a}.
The calculations account for both the phase-speed filters and pupil sizes used for the measurements described in Section~\ref{doug}.   The resulting kernels, ${\bf K}^v$, relate three-dimensional steady flows, $\vel$, to predictions for the EW travel-times differences, 
\begin{equation}
\label{eq.Birch_forward_model}
\delta\tau_{\rm EW}(\br , \Delta) = \int \id^2\br'\id z \; {\bf K}^v(\br'-\br,z ,\Delta) \cdot\vel(\br',z)
\end{equation}
where $\br$ and $\br'$ are two-dimensional position vectors and $z$ is depth.  Both the kernel functions and the flow are vector-valued functions of horizontal position and depth. The travel-time differences are functions of horizontal position and also the pupil size $\Delta$ (notice that for each pupil size there is a corresponding phase-speed filter, see Section~\ref{doug}). 

We assume $\vel$ to be given by the flow field inferred from the inversions of the ridge-filtered travel-time differences shown in Section~\ref{jack}.  We neglect the effects of vertical flows as inversions for depth-dependent vertical flows have not yet been carried out.

\begin{figure}
\includegraphics[width=\textwidth]{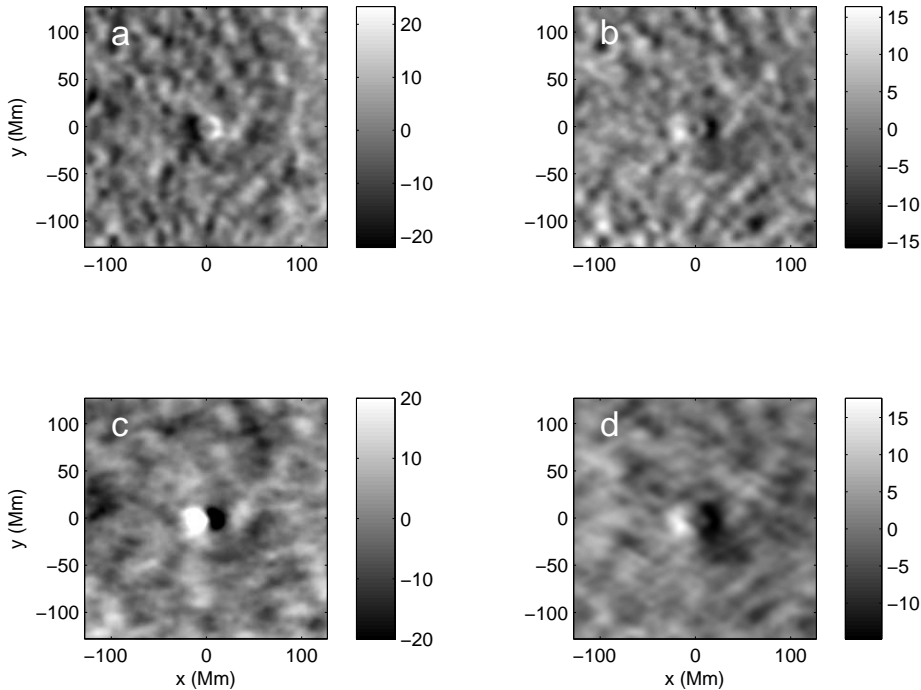}
\caption{\label{Birch_fig1} Measured EW~travel-time differences for phase-speed filters TD1 (panel a), and TD4 (panel c), and the corresponding modeled travel times (panels b and d) for 25 January 2002. The travel-time measurements are described in Section~\ref{doug} and have been smoothed with a Gaussian filter with FWHM of $6.7$~pixels.  The color bars have units of seconds. The gray scales are shown in units of seconds and have been truncated to make the details more visible. The sunspot is centered at roughly $(x,y)=(0,0)$. Notice that the forward model is able to reproduce the moat flow for the filter TD4, but not for the case of TD1.} 
\end{figure}

Figure~\ref{Birch_fig1} compares measurements and forward models of EW travel-time differences for the cases of phase-speed filters TD1 and TD4.  For the case of the filter TD1, the forward model is in qualitative agreement with the measurements in the quiet Sun.  However, the forward model predicts a signature of the moat flow with opposite sign to that seen in the observations.  This shows that the travel-time differences measured using phase-speed filter TD1 and those measured using ridge-filtered travel-times do not yield a consistent picture of the moat flow. Figure~\ref{Birch_fig1} also shows the case of measurements and forward modeling for phase-speed filter TD4.  In this case, and for other filters with large phase speeds, there is qualitative agreement even in the moat.

One possible reason for the disagreements could be that the travel-time sensitivity kernels rely on an imperfect model of the power spectrum of solar oscillations. For example, the model zero-order power spectrum does not include background noise and mode linewidths may not be accurate enough.

We also emphasize that a number of assumptions have been made in carrying out the forward modeling shown in Figure~\ref{Birch_fig1}.  It is known \citep[{e.g.}][]{Gizon2002, Parchevsky2008, Hanasoge2007c} that the reduction in the wave generation rate in sunspots can, in general, produce apparent travel-time differences. For the case of phase-speed filtered travel-time differences this effect has a magnitude of up to 10~s \citep{Hanasoge2007c}.  The magnitude of this effect is not known for travel-times measured using ridge filters.  Similarly, wave damping in sunspots can also produce travel-time differences \citep[{e.g.}][]{Woodard1997, Gizon2002}.  The magnitude of this effect has not been carefully estimated for realistic models of wave absorption in sunspots.  In addition, radiative transfer effects can cause phase shifts in sunpots \citep[{e.g.}][]{Rajaguru2007}.

\subsection{Conflicting Wave-Speed Profiles}

\begin{figure}
\hspace{1.cm}
\includegraphics[width=12cm]{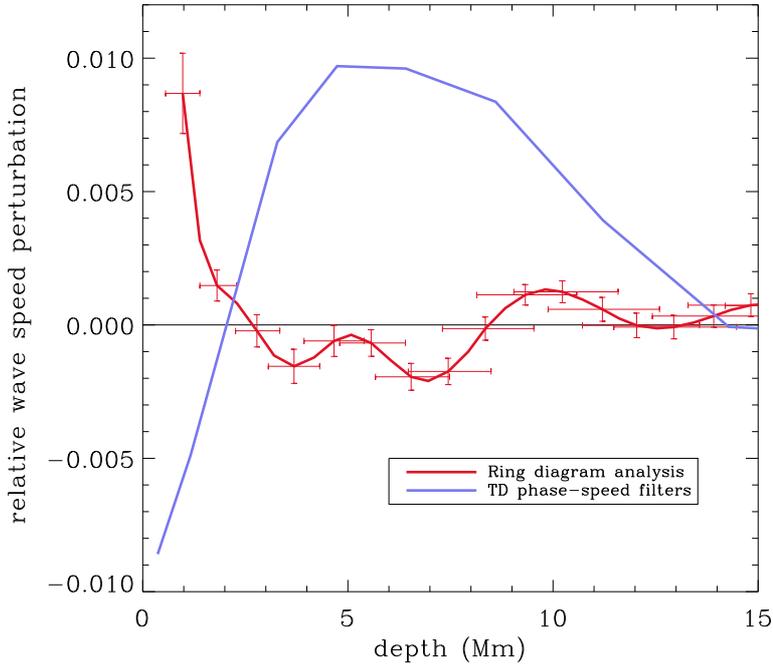}
\caption{Comparison of two different helioseismic methods used to infer wave speed perturbations below AR~9787 ($\delta c_{\rm w}/c$). The red curve shows the averaged ring-diagram results from Figure~\ref{rbcb-1}. The solid blue curve shows the time-distance result (phase-speed filters) from Figure~\ref{fig:avessInversion}, after averaging over the area used for ring analysis. Although they are meant to represent the same quantity, these two curves are noticeably different. }
\label{fig:comp}
\end{figure}

Here we compare the wave-speed inversions from ring analysis (Section~\ref{bogart}) and
from time-distance helioseismology with phase-speed filters (Section~\ref{zharkov}).
To make this comparison possible, we average the TD wave-speed 
inversion over a disk of $15^\circ$ diameter centered on the sunspot, which represents the area used for ring analysis.
The two wave-speed profiles are plotted as a function of depth in Figure~\ref{fig:comp}.
Clearly, they do not match.

How can we explain such a strong disagreement? 
As already mentioned in Sections~\ref{doug} and ~\ref{flowbug}, the  details of the measurement procedures are important for the interpretation of the helioseismic observations; they may not have been fully taken into account in one or possibly both inversions. Although we have not done a TD inversion for wave speed using ridge filters, it is likely that it would give a different answer than the TD inversion using phase-speed filters, thus adding a third curve to Figure~\ref{fig:comp}.  

We also note that both inversions suppose that first-order 
perturbation theory is valid to describe the effect of sunspots on waves.
Unlike the flow perturbation, however, the perturbations in pressure and density introduced
by the sunspot are not small with respect to the quiet-Sun background.
Thus the concept of linear inversions is not necessarily correct for
sunspots and regions of strong magnetic field.
In addition, it is perhaps too naive to model the combined effects of the magnetic 
field in terms of an equivalent sound-speed perturbation.
Ring-diagram inversions do include a contribution from changes
in the first adiabatic exponent, but the direct effect of the magnetic field through the
Lorentz force is not fully accounted for in either inversion.   
We note that the ring-diagram inversions include a treatment of near-surface effects
which is different than in the TD inversions.

\section{Conclusion}
We have studied the sunspot in AR~9787  with several methods of local helioseismology. We have characterized the acoustic wave field near the sunspot and the surrounding plage, measured acoustic absorption by the sunspot, and showed maps of the signature of AR~9787 on the farside of the Sun.
We have shown that the sunspot leaves a strong signature in the observed wave field, as evidenced by strong perturbations in travel times and frequency shifts.  The interpretation of the observations, however, is difficult and we have not been able to draw an unequivocal conclusion about the subsurface structure and dynamics of the sunspot. We have shown that one complication is the extreme sensitivity of helioseismic measurements to the choice of data analysis procedure, such as filtering in frequency-wavenumber space. In addition, our understanding of the effects of strong magnetic fields on solar oscillations is still incomplete. 

On the positive side, we note that the seismically determined moat flow (TD and ridge filters) appears to be consistent with the motion of the MMFs in magnetograms. It is also clear that numerical simulations of wave propagation through model sunspots promise to provide invaluable help in interpreting the observations.

\begin{acknowledgements}
This work work was initiated at the second HELAS local helioseismology workshop held in Freiburg on 7--11 January 2008 and continued during the ISSI Workshop on the Origin and Dynamics of Solar Magnetism held in Bern on 21--25 January 2008. Financial support from project HELAS (European Union FP6), ISSI, project GDC-SDO (Deutsches Zentrum f{\"u}r Luft- und Raumfahrt), and project SISI (European Research Council) is acknow\-ledged.  SOHO is a mission of international cooperation between ESA and NASA. 
\end{acknowledgements}

\bibliographystyle{aps-nameyear}      
\bibliography{seismo_submittedrevision}   

\end{document}